\documentclass[journal=jacsat,manuscript=article]{achemso}
\setkeys{acs}{articletitle = true}

\usepackage[version=3]{mhchem} 
\usepackage{rotating}
\usepackage{tabularx}
\usepackage{threeparttable}
\SectionNumbersOn
\usepackage{xcolor}
\usepackage{adjustbox}
\usepackage{tablefootnote}
\usepackage{lineno}

\author{Danna Qasim}
\affiliation{Solar System Exploration Division, NASA Goddard Space Flight Center, Greenbelt, MD 20771, USA}
\altaffiliation{Department of Physics and Astronomy, Howard University, Washington, DC 20059, USA}
\alsoaffiliation{Center for Research and Exploration in Space Science and Technology, NASA/GSFC, Greenbelt, MD 20771, USA}
\altaffiliation{Current address: Southwest Research Institute, San Antonio, TX 78238, USA}
\email{danna.qasim@nasa.gov}

\author{Hannah L. McLain}
\affiliation{Solar System Exploration Division, NASA Goddard Space Flight Center, Greenbelt, MD 20771, USA}
\altaffiliation{Department of Physics, Catholic University of America, Washington, DC 20064, USA}
\alsoaffiliation{Center for Research and Exploration in Space Science and Technology, NASA/GSFC, Greenbelt, MD 20771, USA}

\author{Jos\'e C. Aponte}
\affiliation{Solar System Exploration Division, NASA Goddard Space Flight Center, Greenbelt, MD 20771, USA}

\author{Daniel P. Glavin}
\affiliation{Solar System Exploration Division, NASA Goddard Space Flight Center, Greenbelt, MD 20771, USA}

\author{Jason P. Dworkin}
\affiliation{Solar System Exploration Division, NASA Goddard Space Flight Center, Greenbelt, MD 20771, USA}

\author{Christopher K. Materese}
\affiliation{Solar System Exploration Division, NASA Goddard Space Flight Center, Greenbelt, MD 20771, USA}
\email{christopher.k.materese@nasa.gov}

\title{Meteorite Parent Body Aqueous Alteration Simulations of Interstellar Residue Analogs}

\keywords{astrochemistry, astrobiology, interstellar ices, asteroids, meteorites, aqueous alteration, radiolysis, liquid chromatography}
\makeatletter
\newcommand*{\forcekeywords}{
  \acs@keywords@print
  \let\acs@keywords@print\relax
}
\makeatother

\begin{document}

\begin{abstract}
 Some families of carbonaceous chondrites are rich in prebiotic organics that may have contributed to the origin of life on Earth and elsewhere. However, the formation and chemical evolution of complex soluble organic molecules from interstellar precursors under relevant parent body conditions has not been thoroughly investigated. In this study, we approach this topic by simulating meteorite parent body aqueous alteration of interstellar residue analogs. The distributions of amines and amino acids are qualitatively and quantitatively investigated, and are linked to closing the gap between interstellar and meteoritic prebiotic organic abundances. We find that the abundance trend of methylamine $>$ ethylamine $>$ glycine $>$ serine $>$ alanine $>$ $\beta$-alanine does not change from pre to post aqueous alteration, suggesting that certain parental cloud conditions have an influential role on the distributions of interstellar-inherited meteoritic organics. However, the abundances for most of the amines and amino acids studied here varied by about twofold when aqueously processed for 7 days at 125$^{\circ}$C, and the changes in the $\alpha$- to $\beta$-alanine ratio were consistent with that of aqueously altered carbonaceous chondrites, pointing to an influential role of meteorite parent body processing on the distributions of interstellar-inherited meteoritic organics. We detected higher abundances of $\alpha$- over $\beta$-alanine, which is opposite to what is typically observed in aqueously altered carbonaceous chondrites; these results may be explained by at least the lack of minerals, inorganic species, and IOM-relevant materials in the experiments. The high abundance of volatile amines in the non-aqueously altered samples suggests that these types of interstellar volatiles can be efficiently transferred to asteroids and comets, supporting the idea of the presence of interstellar organics in solar system objects.  
\end{abstract}

\forcekeywords

\section{Introduction}

The organic inventory of meteorites has been analyzed for decades, as it includes a variety of astrobiologically relevant organic compounds, such as amines and amino acids, that may have contributed to the origin of life on Earth and potentially elsewhere.\citep{glavin2018origin} Substantial efforts have been taken to better understand the origins of these meteorites through direct collection of materials from carbon-rich asteroids, which are thought to be the parent bodies of some carbonaceous chondrites (e.g., OSIRIS-REx and Hayabusa2 sample-return missions). It is important to understand the processes that contributed to the formation and molecular distribution of prebiotic organics in meteoritic and sample-returned materials. Were meteoritic amino acids and amines formed primarily in the parent body, protoplanetary disk, interstellar molecular cloud, or do a combination of these dictate their distribution and abundances?\citep{sandford2020prebiotic,materese2020laboratory} Understanding the origins of organic matter in meteoritic and sample-returned materials aids in elucidating the ubiquity of prebiotic molecules across our solar system and elsewhere.\citep{tachibana2014hayabusa2,lauretta2017osiris,martins2020organic} 

Laboratory experiments involving irradiated interstellar ice analogs have demonstrated the formation of refractory residues, which contain amines, amino acids, and other soluble organic compound classes, suggesting that the prebiotic organic matter found in meteorite parent bodies may have been (at least partially) inherited from interstellar materials.\citep{bernstein2002racemic,munoz2002amino,modica2018amino} Indeed, the work by Martins et al.\citep{martins2015amino} provided convincing evidence that the chemical contents of the Paris meteorite were strongly linked to interstellar precursors. However, the catalog and distribution of amines and amino acids in ice irradiation experiments have not always been consistent with what has been found in carbonaceous chondrites.\citep{elsila2007mechanisms,nuevo2008detailed,aponte2014assessing} For example, some ice irradiation experiments have more $\alpha$- than $\beta$-alanine,\citep{bernstein2002racemic,munoz2002amino} whereas the most aqueously altered chondrites, such as CI1, CM1, and CR1, typically contain more $\beta$- than $\alpha$-alanine, with the ratio reversing with decreasing aqueous alteration in the least altered CM2 and CR2 meteorites.\citep{botta2002relative,modica2018amino,glavin2020abundant} As such, these differences may be accounted for by aqueous alteration of the residues formed from ice irradiation experiments over a range of temperatures. 

Recent laboratory experiments have sought to evaluate the effects of aqueous alteration of commercially available organics that are relevant to the constituents of interstellar materials.\citep{kebukawa2013exploring,kebukawa2017one,vinogradoff2018evolution} Kebukawa et al.\citep{kebukawa2017one} followed a “bottom-up” approach by aqueously altering simple interstellar relevant molecules using calcium hydroxide under different temperature conditions. In contrast, Vinogradoff et al.\citep{vinogradoff2018evolution} used a “top-down” approach by starting with hexamethylenetetramine (HMT), which is an abundant complex photoproduct of interstellar ice chemistry.\citep{bernstein1995organic,cottin2001production,caro2003uv,materese2020production} While these studies provide insight into some of the chemistry that may occur within meteorite parent bodies, there remain gaps in linking the studies to what has been found in aqueously altered meteorites, such as the differences in the $\alpha$-to-$\beta$-alanine ratios. These gaps may be filled by processing of a more representative sample of interstellar-inherited meteoritic compounds, such as interstellar residue analogs.    

This work investigates the effects of meteorite parent body aqueous alteration of interstellar residue analogs on the distribution of the formed amines and amino acids. Although aqueous alteration of the residues of irradiated ices has been performed in the laboratory by Danger et al.,\citep{danger2021exploring} their study simulated the irradiation of ices at the edge of protoplanetary disks, which has higher temperatures and less volatiles available than interstellar ices. Additionally, the work by Danger et al.\citep{danger2021exploring} utilized photons, whereas this work utilizes protons. Albeit, only small differences are found among chemical products of various radiation types. \citep{hudson2001radiation} In this article, we demonstrate the irradiation of a H$_2$O:CO$_2$:CH$_3$OH:$^{15}$NH$_3$ (20:4:2:1) ice mixture at 25 K, and subsequently analyze its products when heated in aqueous solution at 50 and 125$^{\circ}$C for 2, 7 and 30 days. The ice mixture ratio was chosen to reflect the abundances detected towards low-mass young stellar objects (YSOs), where interstellar ices are warmed to temperatures above 20 K.\citep{boogert2015observations} At that stage, ices are exposed to radiation from cosmic rays and the YSO, and residue formation is expected to take place. A fraction of the interstellar ices and organic residues is transferred to the protoplanetary disk,\citep{booth2021inherited,brunken2022major} and become incorporated into meteorite parent bodies, which then undergo aqueous alteration at various temperatures.\citep{zolensky1989aqueous,keil2000thermal,guo2007temperatures} Temperatures of 50 and 125$^{\circ}$C and relatively short heating timescales were chosen for the experiments to reflect the processes in parent bodies that contain CI, CM, and CR chondrites, and may account for the extended periods of time ($\sim$~10$^{4}$ to 10$^{6}$ years) when liquid water was present in the parent body, as they exhibit only moderate degrees of aqueous alteration.\citep{weisberg2006systematics,brearley2006action,elsila2016meteoritic} Additionally, an experimental heating timescale that includes 2 to 30 days has been reported in the literature, \citep{vinogradoff2018evolution,danger2021exploring} and therefore the chemical changes observed in this study can be compared to that of related studies.

The amines and amino acids of interest for this study are methylamine, ethylamine, glycine, alanine, $\beta$-alanine, and serine, as they are the major products formed in these experiments. These compounds have been detected in meteorites,\citep{glavin2008detection,burton2014effects,aponte2020extraterrestrial} and due to their relatively simple structures, may potentially be the more abundant amines and amino acids in interstellar materials. This is supported by the detections of methylamine, ethylamine, and glycine around comet 67P/Churyumov-Gerasimenko\citep{altwegg2016prebiotic} and in samples returned from comet 81P/Wild,\citep{glavin2008detection,elsila2009cometary} and the detections of methylamine in the interstellar medium (ISM).\citep{kaifu1974detection,bogelund2019methylamine} Notably, alanine, $\beta$-alanine, and serine have not been detected in comets or in interstellar environments. Glavin et al.\citep{glavin2008detection} detected L-alanine, $\beta$-alanine, D- and L-serine, and other amino acids in Stardust aerogels exposed to 81P/Wild, but these species can be explained as contamination by their appearance in various controls, thus some or all of these compounds should be viewed as contamination. The near or total homochirality of alanine and serine reported by Glavin et al.\citep{glavin2008detection} indicates that these amino acids are terrestrial contaminants, though as previously noted by Elsila et al.,\citep{elsila2009cometary} it is possible that some of the detected $\beta$-alanine may be cometary. However, the amino acids have been detected in numerous interstellar residue analog experiments, analyzed with and without hydrolysis of the synthesized ice residue.\citep{bernstein2002racemic,munoz2002amino,elsila2007mechanisms} Additionally, the relatively simple distribution of amino acids in the CI1 meteorites Orgueil and Ivuna, dominated by glycine and $\beta$-alanine, was suggested by Ehrenfreund et al.\citep{ehrenfreund2001extraterrestrial} to be evidence that these CI1 meteorites originated from a cometary parent body, and therefore such amino acids are promising components of interstellar-inherited amino acids in meteorites.

\section{Methodology}
\label{method}

\subsection{Synthesis of Residues from Radiation Processing of Ices}

Ices and residues were created within a cryogenic high-vacuum chamber routinely used to investigate proton irradiation of ices (see Hudson and Moore,\citep{hudson1999laboratory} Gerakines and Hudson,\citep{gerakines2013glycine} and Gerakines et al.\citep{gerakines2022radiolytic} for more details). Briefly, the pressures of the chamber at room temperature and just before gas and vapor depositions were $\sim$~$2-3 \times 10^{-7}$ torr and $\sim$~$5-6 \times 10^{-8}$ torr, respectively. A polished aluminum substrate (area $\sim$~5 cm$^{2}$) was cooled to 25 K prior to deposition. On top of the substrate was a small sheet of clean aluminum foil (heated to 500$^{\circ}$C for 24 hours in air) that was compressed to the substrate by a copper gasket. 

Residues were synthesized starting from a H$_2$O:CO$_2$:CH$_3$OH:$^{15}$NH$_3$ (20:4:2:1) ice mixture. Note that $^{15}$NH$_3$ was used to distinguish the results from contamination. Each gas or vapor was placed in its own line within a gas manifold before being simultaneously released into the chamber through calibrated leak valves. Leak valve calibrations and laser interferometry were jointly used; each leak valve was calibrated to a value that corresponded to a specific ice growth rate needed to acquire the desired ice mixture ratio. The deposition rate of each molecular species on the foil was monitored by laser interferometry. To determine the leak valve value that corresponded to the needed deposition rate, a large range of leak valve calibrations was performed to create calibration curves for three of the molecules (arbitrarily chosen), with the deposition rate of the remaining molecule as the independent variable. With laser interference data and data from the literature (Table~\ref{table1}), the following equation was used to determine the desired deposition rates to obtain a 20:4:2:1 ice mixture: 

\begin{equation}\label{eqn1}
\frac{\text{molecule X deposition rate}}{\text{molecule Y deposition rate}} = \frac{\text{N}_{\mathrm{molecule\;X}}}{\text{N}_{\mathrm{molecule\;Y}}} \times \frac{\text{$\rho$}_{\mathrm{molecule\;Y}}}{\text{$\rho$}_{\mathrm{molecule\;X}}} \times \frac{\text{M}_{\mathrm{molecule\;X}}}{\text{M}_{\mathrm{molecule\;Y}}} \times \frac{\textit{n}_{\mathrm{molecule\;X}}}{\textit{n}_{\mathrm{molecule\;Y}}}
\end{equation}

\

\noindent where X and Y are arbitrary molecules, N = number of molecules (or moles), $\rho$ = density (g cm$^{-3}$), M = molar mass (g mol$^{-1}$), and \textit{n} = refractive index. Ices were grown for three hours to reach a thickness of $\sim$~15 $\mu$m, which the ice thickness can be determined by the following equation: 

\begin{equation}\label{eqn3}
\textit{h} = \frac{\textit{N}_\text{\text{f}r}\lambda}{\text{2}\sqrt{\textit{n}^{2}-\sin^{2}{\theta}}}
\end{equation}

\

\noindent where \textit{h} = ice thickness, \textit{N}$_{\text{f}\text{r}}$ = number of fringes, $\lambda$ = wavelength of He-Ne laser (670 nm), and $\theta$ is the is the incidence angle of the laser beam.\citep{heavens1991optical} A thickness of $\sim$~15 $\mu$m was chosen, as it resulted in a detectable amount of product, and is also below the penetration depth of $\sim$~25 $\mu$m for 0.9 MeV protons.  

\begin{table*}
	\centering
	\caption{Values used to calculate the deposition rate for each molecule to obtain the chosen ice mixture ratio.}
    \label{table1}
	\begin{tabular}{c c c c } 
		\hline
        Molecule (Brand and Purity) & \textit{n} & $\rho$ & References for \textit{n} and $\rho$
		\\
 
		\hline
        
		H$_2$O (Fisher Chemical HPLC Grade) & 1.31 & 0.94 & Weast et al.\citep{weast1984handbook} Narten et al.\citep{narten1976diffraction}\\ 
		CO$_2$ (Matheson 99.8\%) & 1.33 & 1.40 & Loeffler et al.\citep{loeffler2016effects}\\
        CH$_3$OH (Sigma-Aldrich HPLC Grade) & 1.26 & 0.64 & Luna et al.\citep{luna2018densities} \\ 
		$^{15}$NH$_3$ (Cambridge Isotope Laboratories 98\%) & 1.41$^{a}$ & 0.78$^{b}$ & - \\ 
 
\hline
\end{tabular}
\begin{tablenotes}
\item \small $^{a}$ denotes that n is the value for $^{14}$NH$_3$ from Bouilloud et al.\citep{bouilloud2015bibliographic} $^{b}$ denotes that the value is estimated by multiplying $\rho$ of $^{14}$NH$_3$ (obtained from Bouilloud et al.\citep{bouilloud2015bibliographic}) by the molar mass ratio of \( \mathrm{\frac{^{15}NH_3}{^{14}NH_3}} \).
\end{tablenotes}
\end{table*}

After the ices were formed, protons with an energy of 0.9 MeV at a current of $1.5 \times 10^{-7}$A were admitted into the vacuum chamber through an interfaced beamline that was connected to a Van de Graaff accelerator. Infrared (IR) spectra of the ices and residues were obtained \textit{in situ} with a Nicolet Nexus 670 spectrometer. Ice and residue spectra were recorded as the average of 100 scans, with a resolution of 2 cm$^{-1}$, from 5000 to 650 cm$^{-1}$. Although IR spectroscopy was not used as the primary analytical technique, its purpose was to ensure consistency of the infrared features between repeated experiments, as well as to check for complex molecular structures in the residue. The ices experienced a radiation dose of 10 eV/molecule, which is roughly comparable to the dose received by interstellar icy grains in a dense cloud over the span of 10$^{7}$ years.\citep{moore2001mid} The radiation dose was calculated by the following equation: 

\begin{equation}\label{eqn5}
\text{Dose (MGy)} = \text{SF} \times (1.602 \times 10^{-22})
\end{equation}

\noindent where S is the proton stopping power (eV cm$^{2}$ g$^{-1}$ p$^{+-1}$), F is the proton fluence (p$^{+}$ cm$^{-2}$) that is determined from the integrated current through the aluminum substrate sample, and the factor of $1.602 \times 10^{-22}$ converts radiation dose units from eV g$^{-1}$ to MGy. The stopping power value of $2.583 \times 10^{8}$ eV cm$^{2}$ g$^{-1}$ p$^{+-1}$ was determined from the software package, Stopping and Range of Ions in Matter.\citep{ziegler2010srim}  

After irradiation, the samples were warmed to 300 K at 1.5 K/min to slowly release volatiles while keeping the residue intact within the foil. The cryostat was pulled out of the vacuum chamber, and the residue-covered foils were pulled out by baked tweezers and carefully placed into $13 \times 110$ mm baked tubes (Kimble). The tubes were then immediately stored in a -80$^{\circ}$C freezer prior to preparation for the aqueous alteration experiments, as it was most effective to have a bulk of samples created before proceeding to the next steps. Thus, samples were frequently stored in the -80$^{\circ}$C freezer throughout the entire experimental procedure. As the samples are eventually processed, storage of the samples at -80$^{\circ}$C would have a negligible effect on the overall results.       

\subsection{Preparation of Residues and Controls for Aqueous Alteration}

All glassware was cleaned by rinsing with Milli-Q ultrapure water (18.2 M$\Omega$, $<$ 3 ppb total organic carbon), then baked in a muffle furnace at 500$^{\circ}$C in air overnight. To extract the samples from the foils, 850 $\mu$L of methanol (Fisher Chemical HPLC Grade) were pipetted into each of the foil-containing tubes, and the tubes were subsequently sonicated for 10 minutes. After sonication, the resulting liquids were pipetted into $12 \times 32$ mm baked vials (Supelco) that contained 10 $\mu$L of 6M HCl (Tama Chemicals Co., Ltd ultrapure), used to avoid the loss of volatile amines. The acidified methanol solutions were split into four 200 $\mu$L fractions placed in clean $13 \times 100$ mm borosilicate tubes (Kimble). One tube was designated to not be aqueously processed, another tube designated to be processed at 50$^{\circ}$C, another tube designated to be processed at 125$^{\circ}$C, and the last tube held in reserve as a spare. This process was repeated in triplicates to account for the three different processing time periods we studied (2, 7, and 30 days). Following this procedure, the four tubes were placed into a Labconco CentriVap for $\sim$~24 hours to remove the methanol. The dried tubes were then removed, and 200 $\mu$L of ultrapure water were added to each of the samples. The pH of each water-filled tube was measured by spotting a drop of solution on universal pH paper to check for pH drift from $\sim$~7-10 over the course of the experiment since the samples were unbuffered. The necks of the tubes were formed at approximately the top 1/3 of the tube with an oxy-propane torch to create an ampoule. The torch was affixed to a bracket, and the tube was manually rotated while held at both ends with nitrile-gloved hands to prevent potential sample contamination. \citep{dworkin2018osiris} While the necks were formed, the solutions were not permitted to warm significantly, as verified by monitoring the glass tube by the sample while manually sealing by touch. The ampoules were liquid nitrogen freeze-pumped-thaw degassed three times to remove volatile air contaminants, then flame sealed \textit{in vacuo} as described in Pavlov et al.\citep{pavlov2022rapid}. 

\subsection{Aqueous Alteration}

The aqueous alteration samples were processed for a given temperature (50 or 125$^{\circ}$C) and duration (2, 7, or 30 days). To heat the samples, the tubes were placed within dry bath blocks, which were placed within a furnace or oven (Thermo Scientific). The tubes were processed uninterrupted during their 2, 7, or 30 days duration. Note that the non-processed samples were placed in the freezer to preserve the products.     

\subsection{Preparation of Aqueously Altered Samples for Analysis}

After aqueous alteration of the samples, the ampoule seals were carefully broken, and the pH was measured and compared to the non-processed sample. All samples had a pH of $\sim$~7 except the sample processed at 125$^{\circ}$C for 30 days, which drifted to pH $\sim$~8  (this pH range is consistent with what has been modeled for CM asteroid fluids during aqueous alteration, pH $\sim$~7-10).\citep{zolensky1989aqueous} To obtain most of the aqueously altered samples, 400 $\mu$L of ultrapure water were pipetted into the tubes. The liquids in the tubes were transferred into baked $12 \times 32$ mm Total Recovery Autosampler Vials (Waters) that contained 20 $\mu$L of 6M HCl and dried in a centrifugal evaporator. 

Samples were derivatized with AccQ•Tag reagents according to the Waters manufacturer’s protocol. 80$\mu$L of Waters AccQ•Tag sodium borate buffer and then 20 $\mu$L of Waters AccQ•Tag derivatization agent was added. Samples and standards were heated for 10 minutes at 55$^{\circ}$C immediately following the addition of the derivatizing agent. Standards consisting of a set of 9 calibrators (0.25 to 250 $\mu$M)  of the AccQ•Tag Amino Acid Standard (purity $\geq$96.8\%) along with \textit{$\gamma$}-aminobutyric acid ($\gamma$-ABA), D,L-$\beta$-aminoisobutyric acid ($\beta$-AIB), D,L-$\beta$-amino-\textit{n}-butyric acid ($\beta$-ABA), D,L-$\alpha$-aminoisobutyric acid ($\alpha$-AIB), and D,L-$\alpha$-amino-\textit{n}-butyric acid ($\alpha$-ABA) were prepared in water and treated the same way. The UHPLC solvents were prepared according to the Waters manufacturer’s protocol.

\subsection{Analysis}

Separation of amino acids was accomplished by injecting 1 $\mu$L of the AccQ•Tag derivatized sample onto an Acquity AccQ•Tag Ultra C18 column, $150 \times 2.1$ mm column (1.7 $\mu$m particle size) maintained at 55$^{\circ}$C. Chromatographic separation was achieved using 100 $\mu$L AccQ•Tag concentrate A with 900 $\mu$L of ultrapure water as eluent A and Waters AccQ•Tag B as eluent B. Analytes were eluted using a flow rate of 700 $\mu$L/min and the following gradient time in minutes (\% B): 0.54 (0.1), 5.74 (10), 7.74 (21.2), 8.04 (59.6), 8.64 (59.6), 8.73 (0.1), 10.00 (0.1). The Waters Acquity UHPLC was equipped with a fluorescence detector set to $\lambda$$_{\textrm{excitation}}$ = 266 nm and $\lambda$$_{\textrm{emission}}$ = 473 nm. 

AccQ•Tag measurements were conducted using a Xevo G2 XS with the electrospray capillary voltage set to 1.2 kV, the sample cone to 40 V, the source temperature to 120$^{\circ}$C, the cone gas flow to 70 L hr$^{-1}$, the desolvation temperature to 500$^{\circ}$C, and the desolvation gas flow to 1000 L/h. The ToF-MS analyzer was operated in V-optics mode, which used a reflectron to achieve a full width at half maximum resolution of 22,000 based on the m/z value of leucine enkaphalin. The m/z range over which data were collected was 100-600. For UHPLC analysis, a 100 $\mu$L syringe and 15 $\mu$L needle were used.

A set of 23 individual amino acids were prepared in water and analyzed at 9 different concentrations to generate a linear least-square model fit for each analyte (Supporting Information). A set of 16 individual amines were prepared in water and analyzed at 8 different concentrations to generate a linear least-square model for each amine analyte (Supporting Information). The abundance of amines and amino acids were quantified from peak areas generated from the mass chromatogram of their AccQ•Tag derivatives. The abundances are defined as the average of three separate measurements of the same extracted sample.

No unexpected or unusually high safety hazards were encountered.

\section{Results}
\label{results}

\begin{figure}
\includegraphics[width=8cm]{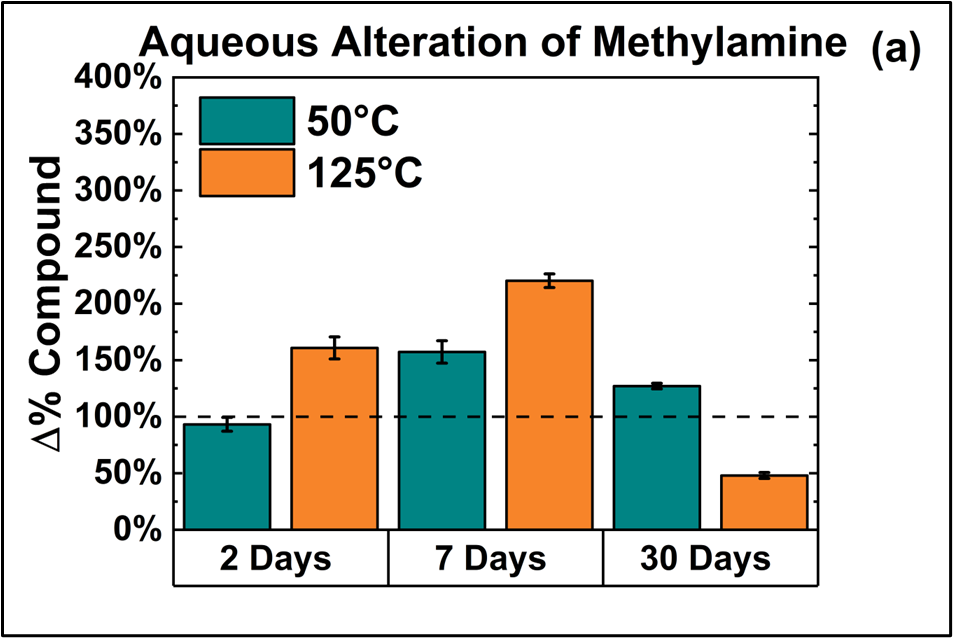}
\includegraphics[width=8cm]{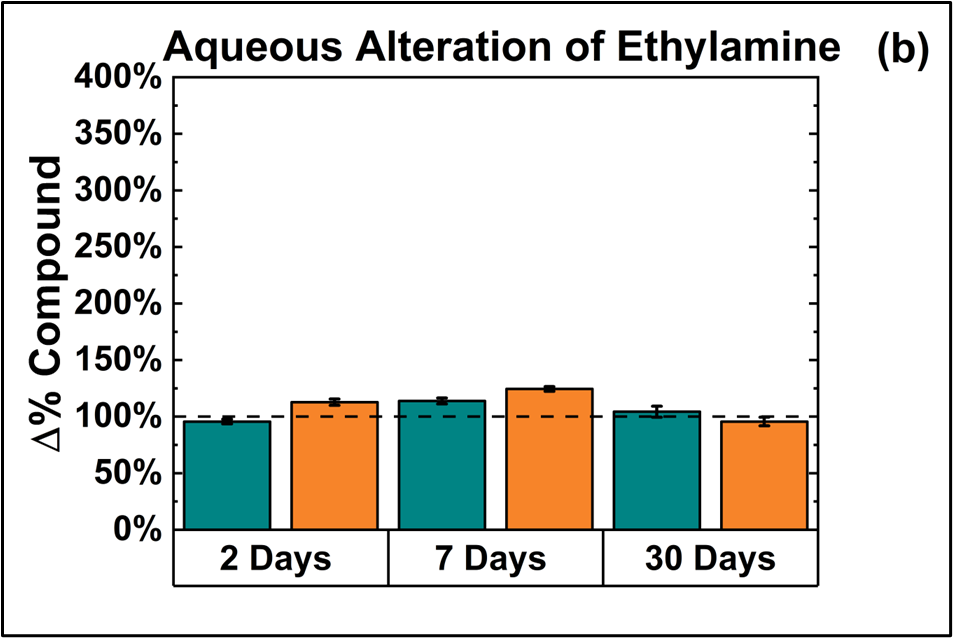}
\includegraphics[width=8cm]{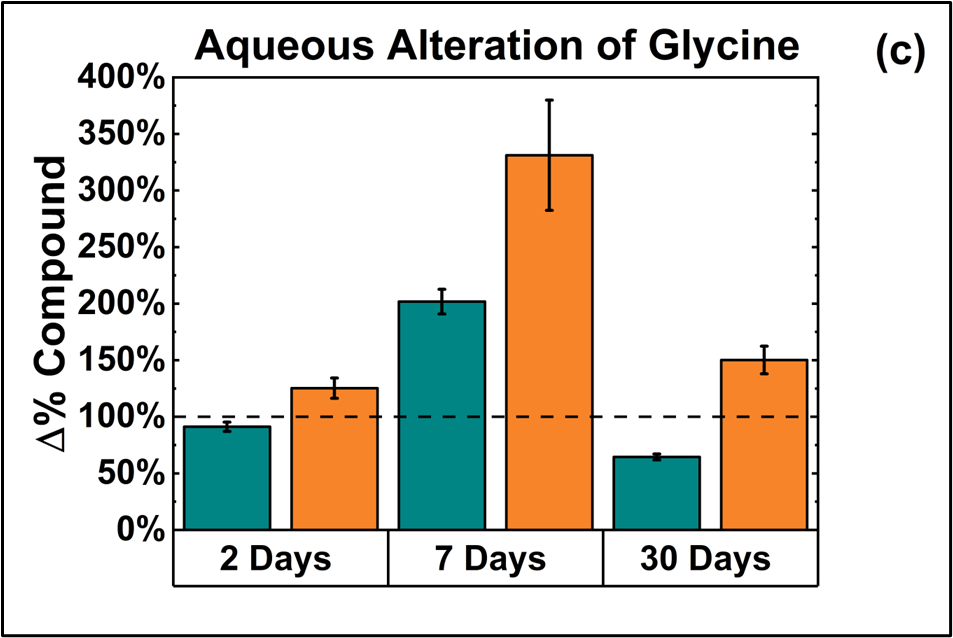}
\includegraphics[width=8cm]{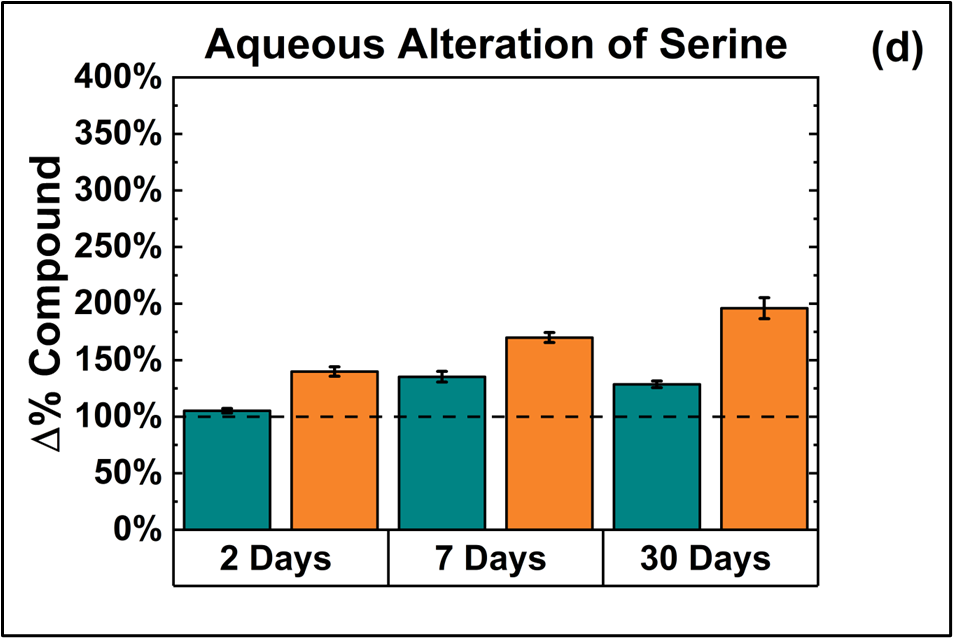}
\includegraphics[width=8cm]{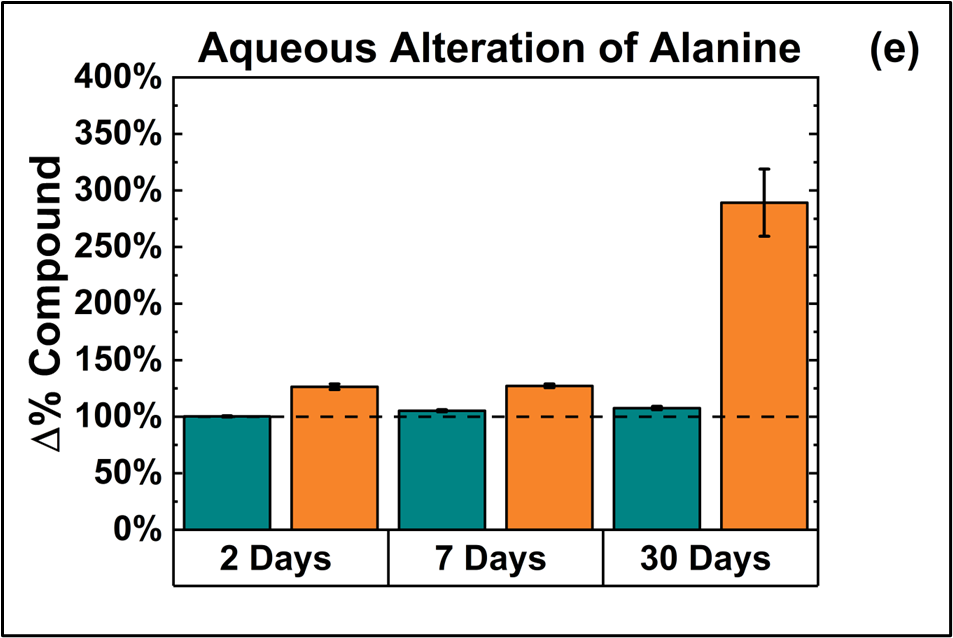}
\includegraphics[width=8cm]{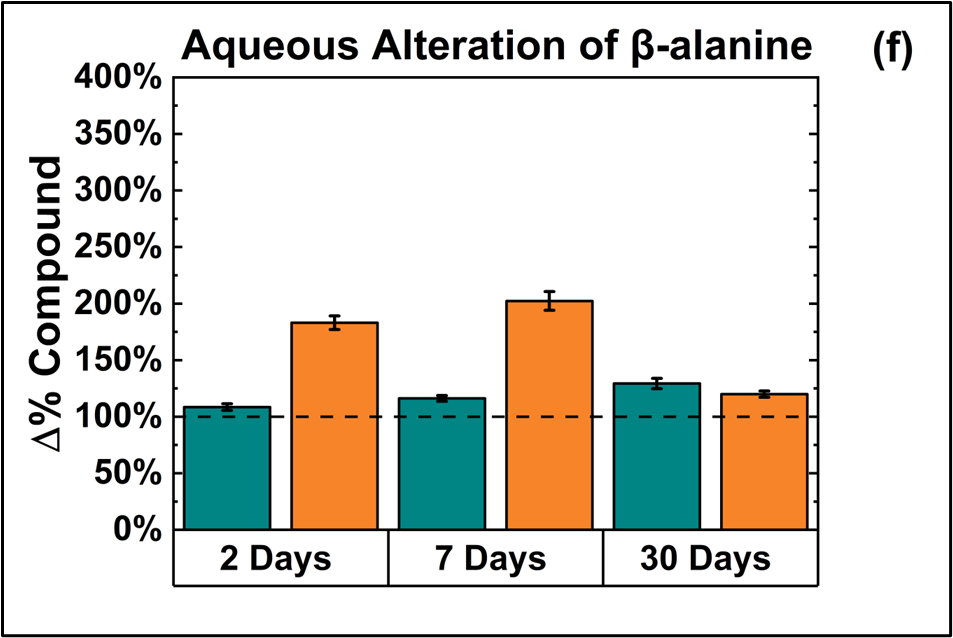}
\caption{Change in product abundances after aqueous alteration of the residues for (a) methylamine, (b) ethylamine, (c) glycine, (d) serine, (e) alanine, and (f) $\beta$-alanine. \textit{Caution should be taken when analyzing the 30 days duration, 125$^{\circ}$C sample (see text)}. Columns below the dashed line indicate a decrease in abundance after aqueous alteration, and columns above the dashed line indicate an increase in abundance. The aqueously altered values are normalized to the non-aqueously altered values. Note that because each duration represents a different residue, \textit{comparisons can only be made between data reflecting the same durations and between the aqueously altered and non-aqueously altered values}.}   
\label{fig1}
\end{figure}

The major product abundances are shown in Table~\ref{table2}. Minor product abundances include $\gamma$-aminobutyric acid, $\beta$-aminoisobutyric acid, isopropylamine, and propylamine (Supporting Information). Residues A, B, and C were formed by three separate, identical ice irradiations. These residues were either not processed (i.e., initial) or separately aqueously altered at temperatures of 50 and 125$^{\circ}$C, for durations of 2 days (residue A), 7 days (residue B), and 30 days (residue C). As each residue was formed on a different date, each one was exposed to unique conditions that contributed to variations in the starting abundances. Figure~\ref{fig1} shows the $\bigtriangleup$\% in product abundances for (a) methylamine, (b) ethylamine, (c) glycine, (d) serine, (e) alanine, and (f) $\beta$-alanine as a function of the aqueous alteration duration and temperature in order to display aqueous alteration trends. Note that the aqueously altered values are normalized to the non-aqueously altered values.

\begin{table*}
\centering
\begin{adjustbox}{width=1\textwidth}
	\caption{Product abundances of the aqueously altered interstellar residue analogs. Residues A, B, and C represent the same residue formed on different dates, hence having different initial values. Amino acid concentrations were determined from both UV fluorescence and single ion mass peak areas and included background level correction using a procedural blank and a comparison of the peak areas with those of an amino acid standard run on the same day. The reported uncertainties ($\delta$x) are based on the standard deviation of the average value of three separate measurements (n) with a standard error, $\delta$x = $\sigma$x · (n)$^{1/2}$ and take into account peak resolution quality.}
	\label{table2}
	\begin{tabular}{c c|| c |c c c| c c c} 
		\hline
  \multicolumn{9}{c}{Abundances (nmol g$^{-1}$)}\\
  \hline
        Compound & Residue & Initial & 2 Days & 7 Days & 30 Days & 2 Days & 7 Days & 30 Days\\
& & & & (50$^{\circ}$C) & & & (125$^{\circ}$C) &  \\ 
\hline
 & A & 2230$\pm$130 &2080$\pm$190  &- &- &3590$\pm$390 &- &-\\
Methylamine & B & 900$\pm$51 & - & 1410$\pm$160& -& -&1970$\pm$100 &-\\
 & C & 1920$\pm$210 & - &- & 2440$\pm$280& -&- &923$\pm$57\\
 \hline
& A & 880$\pm$29& 842$\pm$40& -&- & 992$\pm$10&- &-\\
Ethylamine & B & 663$\pm$8& -& 754$\pm$25&- & -&824$\pm$23 &- \\
& C & 826$\pm$54& -& -&862$\pm$23 &- &- & 790$\pm$25\\
\hline
& A & 47.5$\pm$3.7& 43.3$\pm$3.3&- &- &59.56$\pm$0.51 &- &- \\
Glycine & B & 14.8$\pm$1.5& -& 29.8$\pm$4.5&- & -&48.9$\pm$2.9 &- \\
& C &64.5$\pm$7.6 &- & -& 41.5$\pm$4.5&- &- &96.9$\pm$9.7 \\
\hline
 & A &6.10$\pm$0.18 & 6.42$\pm$0.29&-&- &8.53$\pm$0.50 &- &- \\
Serine & B &4.96$\pm$0.13 &-& 6.71$\pm$0.39&- & -& 8.43$\pm$0.2&- \\
 & C &6.39$\pm$0.27 &- &- &8.22$\pm$0.52 &- &- &12.52$\pm$0.89 \\
 \hline
 & A &3.82$\pm$0.02 &3.83$\pm$0.03 &- &- &4.83$\pm$0.10 &- &- \\
Alanine & B &3.63$\pm$0.02 &- &3.82$\pm$0.05 &- &- &4.62$\pm$0.07 &- \\
 & C &3.81$\pm$0.02 & -&- &4.10$\pm$0.07 & -&- &11.0$\pm$1.1 \\
 \hline
 & A &1.89$\pm$0.01 &2.05$\pm$0.05 &- &- &3.46$\pm$0.12 & -&- \\
$\beta$-alanine & B &1.73$\pm$0.01 & -& 2.01$\pm$0.05 &-&- &3.50$\pm$0.17&- \\
 & C &1.91$\pm$0.03 &- &- & 2.47$\pm$0.12&- &- &2.29$\pm$0.09 \\

\hline
\end{tabular}
\end{adjustbox}
\end{table*}

\clearpage

\section{Discussion}
\label{discussion}

The non-aqueously altered abundances of amines and amino acids (i.e., samples from solely ice irradiation) varied between residues A, B, and C by up to $\sim$~4×. However, product abundances followed the same trend, methylamine $>$ ethylamine $>$ glycine $>$ serine $>$ alanine $>$ $\beta$-alanine in all samples, including those for aqueously altered samples for all temperature and duration combinations (Table~\ref{table2}). These findings suggest that the set of conditions for forming the ice residues have a dominant role in the total concentrations of amines and amino acids formed pre-aqueous alteration, but because the abundance of the formed compounds decreased with increasing molecular weight and because amines were found in higher concentrations than amino acids in all residues (even after hydrothermal processing), we found that slight variations of the conditions we used to synthesize the interstellar residue analogs do not alter the absolute molecular distributions of amines and amino acids made.    

The impact of aqueous alteration on the amine and amino acid distributions varied, depending on the alteration temperature and duration, as well as the molecule. As shown in Figure~\ref{fig1}, for durations of 2 and 7 days, the abundances were highest at 125$^{\circ}$C for all compounds studied. 2 days of processing at 50$^{\circ}$C negligibly impacted the abundances. It is distinctly shown in the 7 days duration that the abundances increased in a stepwise fashion (i.e., non-aqueously altered $<$ 50$^{\circ}$C $<$ 125$^{\circ}$C). The 30 days, 125$^{\circ}$C sample showed a modest pH drift from 7 to 8. It is possible that this was due to borate leaching from the glass or unknown contamination or a procedural error. This makes the data suspect and difficult to interpret the data points. However, observation of the 30 days, 50$^{\circ}$C data in comparison to the non-aqueously altered values shows no clear aqueous alteration trends. For the overall trend, ethylamine and alanine were the least impacted by aqueous alteration. The spike in the alanine abundance shown in Figure~\ref{fig1} appears to be an anomaly due to the potentially contaminated sample. 

\begin{figure}
\includegraphics[width=\textwidth]{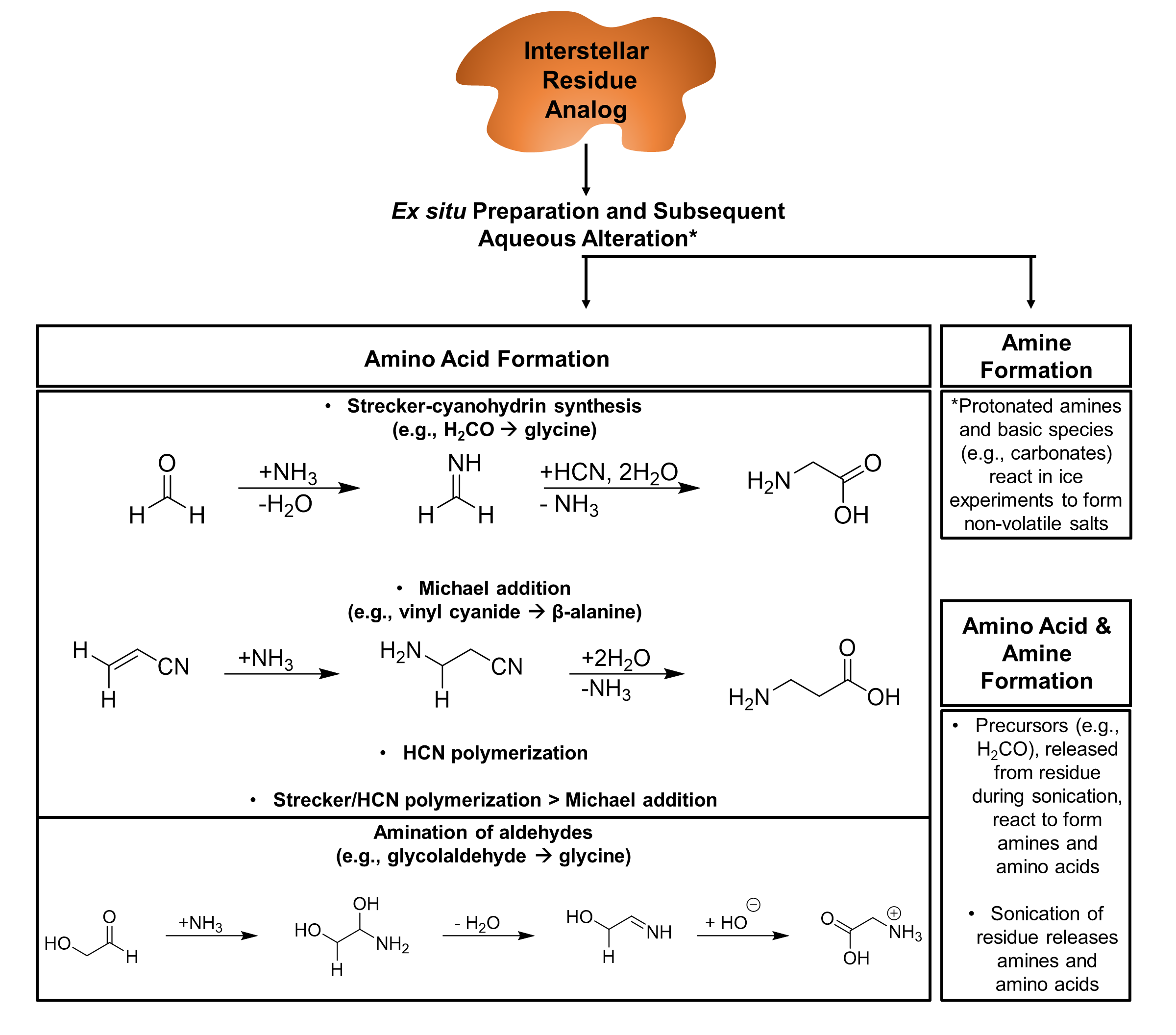}
\caption{An illustration summarizing how the amines and amino acids detected after aqueous alteration of the interstellar residue analogs were formed. Amine-containing salts were formed solely in the ice experiments, as noted by *.}   
\label{fig1.5}
\end{figure}

The observed aqueous alteration trends can be partially interpreted through consideration of synthesis routes. An illustration summarizing the synthesis routes, as well as other pathways to forming the detected amines and amino acids, is shown in Figure~\ref{fig1.5}. Amines and amino acids could have been formed during the aqueous alteration process through precursors that were freed from the residue by the sonication extraction step. This is similar to what is observed during acid hydrolysis of meteorite extracts, where larger abundances of amines and amino acids are returned from the partial breakup of insoluble organic matter (IOM), and other structurally related species.\citep{simkus2019methodologies} 

From our aqueous alteration experiments, we did not observe a clear trend of consumption and production of amines and amino acids sharing the same aliphatic backbone when comparing data of the same durations for any of the experiments. For example, the abundance of ethylamine increases from $\sim$~842 to 992 nmol g$^{-1}$ when going from 50 to 125$^{\circ}$C for the 2 days duration, and the abundance trend of $\beta$-alanine also shows an increase. Therefore, it was determined that the amine and amino acid products found here did not share parent-daughter synthetic relationships.

The abundances of methylamine and ethylamine were at least an order of magnitude higher than the amino acid abundances in the non-aqueously altered samples. As both amines would desorb below room temperature in an ice experiment\citep{forstel2017formation} and are volatile in ambient air, their substantial presence in the non-aqueously altered samples would occur if they were trapped by and subsequently released from the ice residue, and/or synthesized from precursor molecules in the extraction step. It is also likely that basic species like carbonates may have formed during ice irradiation,\citep{moore1991infrared} species which may have reacted with protonated methylamine and ethylamine to form non-volatile salts and thus, rendering protection from volatilization after irradiation.

Synthesis routes that include nitriles may have a major role in influencing the aqueous alteration trends. Some key mechanisms, such as the Strecker-cyanohydrin synthesis, the Michael addition to $\alpha$-$\beta$-nitriles, and the polymerization of HCN, dictate the formation of $\alpha$- and $\beta$-amino acids in aqueously altered environments, in which they include species like cyanide (CN–) and aliphatic nitriles  (e.g., acrylonitrile).\citep{simkus2019methodologies} In our experiments, cyanide and aliphatic nitriles could have formed in the ice starting from $^{15}$NH$_3$, as amines can be converted to nitriles through radiation processes.\citep{palumbo2000roc,danger2011experimental} Indeed, cyanide was found in the spare unprocessed samples. Interestingly, since CN– formed from the ice irradiation process, one would expect that they would participate in a number of chemical routes during aqueous alteration, and therefore potentially change the product abundance trend. A possible explanation is that under our experimental conditions, the efficiencies of the mechanisms that include cyanides or aliphatic nitriles changed similarly between each other under different durations and temperatures. However, that is difficult to determine considering that there is a variety of energy barrier values for each mechanism.\citep{pardo1993mechanisms,magrino2021step} Due to the higher abundances of glycine and $\alpha$-alanine than $\beta$-alanine, it is proposed that Michael addition is a less dominant mechanism than Strecker/HCN polymerization reactions.   

The observed aqueous alteration trends can also be interpreted through analysis of a potential degradation process of complex molecular structures formed during ice irradiation. Signatures of refractory organics and polymers were searched for in the IR data of the irradiated ice (25 K) and warmed residue (300 K). Species such as hexamethylenetetramine (HMT), along with variants and polymethylenimine (PMI), have been identified in the IR data of residues formed from the photolysis and irradiation of ices.\citep{briggs1992comet,bernstein1995organic,cottin2001production,vinogradoff2013importance,materese2020production} The identification of such molecules in the IR spectra of complex ices is not trivial and we could not definitively identify any of the listed species in our samples. Regardless, the presence of refractory organics and polymers in the literature involving the irradiation of similar precursor ices makes their presence in our experiments likely.    

As mentioned previously, an aqueous alteration trend was not observed for 30 days, suggesting a depletion of the initial precursor species at 30 days. However, it could also reflect the depletion of complex molecular structures in the aqueously altered environment that were sources of precursor molecules and/or the amines and amino acids themselves. In a related study by Vinogradoff et al.,\citep{vinogradoff2018evolution} it was observed that HMT did not persist after 7 days of aqueous alteration at 150$^{\circ}$C, and that the molecular size index and the aromatic/olefine/imine index dropped after 20 days of processing. Therefore, in our experiments, it is possible that polymers and/or other large molecular structures that acted as sources of precursor molecules and/or the products substantially depleted by 30 days of aqueous alteration. Note that thermal degradation was not considered, as the amino acids investigated in this study only start to decompose in the solid-state at temperatures well above those used in this study.\citep{weiss2018thermal,rodante1992thermodynamics}

Our experiments showed an overabundance of glycine to alanine, as well as the presence of $\beta$-alanine. In the work by Vinogradoff et al.,\citep{vinogradoff2020impact} the formose reaction, followed by the addition of NH$_3$, yielded glycine. The addition of H$_2$CO to glycine then yielded alanine, thus, alanine abundances were less than that of glycine. The presence of $\beta$-alanine originated from the addition of NH$_3$ to glycolaldehyde formed in the formose reaction. However, unlike the study by Vinogradoff et al.,\citep{vinogradoff2020impact} our experiments were performed under neutral pH conditions (and with a potentially limited amount of H$_2$CO), whereas the hydrothermal formose reaction is well known to be base and divalent cation catalyzed.\citep{breslow1959mechanism,kopetzki2011hydrothermal} Hence, the formose reaction is not expected to be a dominant reaction in our experiments. However, the slow step in the formose reaction is the formation of glycolaldehyde. If sufficient aldehydes were formed during the ice irradiation, as proposed to occur in irradiated methanol-containing ices\citep{hudson2005ir} and observed to form from laboratory astrophysical ices,\citep{de2015aldehydes,fedoseev2015experimental} it is possible that the amination could have occurred to these aldehydes in solution. There is some evidence that this mechanism could have been a minor component due to the formation of small amounts of $\gamma$-aminobutyric acid, which is not generated by Strecker or Michael addition, but was observed by Vinogradoff et al.\citep{vinogradoff2020impact}

\section{Connection between interstellar and meteorite parent body organics}

With and without aqueous alteration, the abundance trend of methylamine $>$ ethylamine $>$ glycine $>$ serine $>$ alanine $>$ $\beta$-alanine did not change. Thus, the results from this study imply that interstellar cloud conditions and chemical inventories may have a highly influential role in the abundance trend of amines and amino acids in meteorite parent bodies, even after the parent body has been aqueously altered. These results parallel meteoritic studies, such as the work by Martins et al.\citep{martins2015amino}, that have demonstrated the strong connection between meteoritic and interstellar organic content. Additionally, the overall trend did not change even though the non-aqueously altered abundances were different for each residue, and cyanide was present, which is a significant reactant in aqueous alteration mechanisms. The resilience of the trend to the aqueous alteration and initial conditions of our experiments suggests that for parent bodies experiencing aqueous alteration at temperatures $\leq$ 125$^{\circ}$C, the amine and amino acid abundance trend can be somewhat linked to that from the parent molecular cloud, assuming the molecules were interstellar-inherited. Albeit, a major caveat to this scenario is the absence of IOM-relevant materials, minerals and iron, magnesium, aluminum, and other inorganic species with varying oxidation states, which were not used in our experiments, and may be pivotal for the synthesis (and/or destruction) of chondritic amines and amino acids. However, it should be noted that dissimilarities in amino acid distributions have been found between CI-like chondrites and sample-returned materials.\citep{parker2022amino} Thus, progress is needed in both fields of study to better understand the interstellar-meteorite parent body connection.

The aqueous alteration abundances of ethylamine and alanine remained close to that of the non-aqueous alteration abundances, also suggesting that the abundances of these molecules post aqueous alteration may reflect the abundances delivered from the interstellar parental cloud. A major caveat includes the impact that parent body minerals would have on the abundances during aqueous alteration. Alternatively, methylamine, glycine, serine, and $\beta$-alanine abundances were altered by $\sim$~twofold. Thus, when analyzing the abundances in meteorites, it should be considered that the abundances of those molecules are likely not reflective of the distributions solely from interstellar clouds.      

Similar to Modica et al.,\citep{modica2018amino} we found higher abundances of $\alpha$- than $\beta$-alanine in our synthetic interstellar ice residues. As mentioned previously, the most aqueously altered meteorites typically exhibit higher abundances of $\beta$- relative to $\alpha$-alanine.\citep{botta2002relative,modica2018amino,glavin2020abundant} In the aqueous alteration studies by Kebukawa et al.\citep{kebukawa2017one} and Vinogradoff et al.,\citep{vinogradoff2020impact} only ammonia or an amine was utilized as the N-bearing reactant, and in all cases, there was a greater abundance of $\alpha$- than $\beta$-alanine. A phyllosilicate was incorporated in the study by Vinogradoff et al.,\citep{vinogradoff2020impact} which still lead to more $\alpha$- than $\beta$-alanine, albeit cyanide and aliphatic nitriles were not used as reactants. These findings suggest that cyanide and aliphatic nitriles reacting with minerals in the parent body may be an influential piece of the puzzle in connecting interstellar abundances to the amine and amino acid abundances found in aqueously altered chondrites. Future experimental analyses including representative meteoritic matrices and unoxidized mineral species are needed to fully understand the molecular distribution and abundance evolution during parent body aqueous alteration.       

\begin{figure}[H]
\centering
\includegraphics[width=\textwidth]{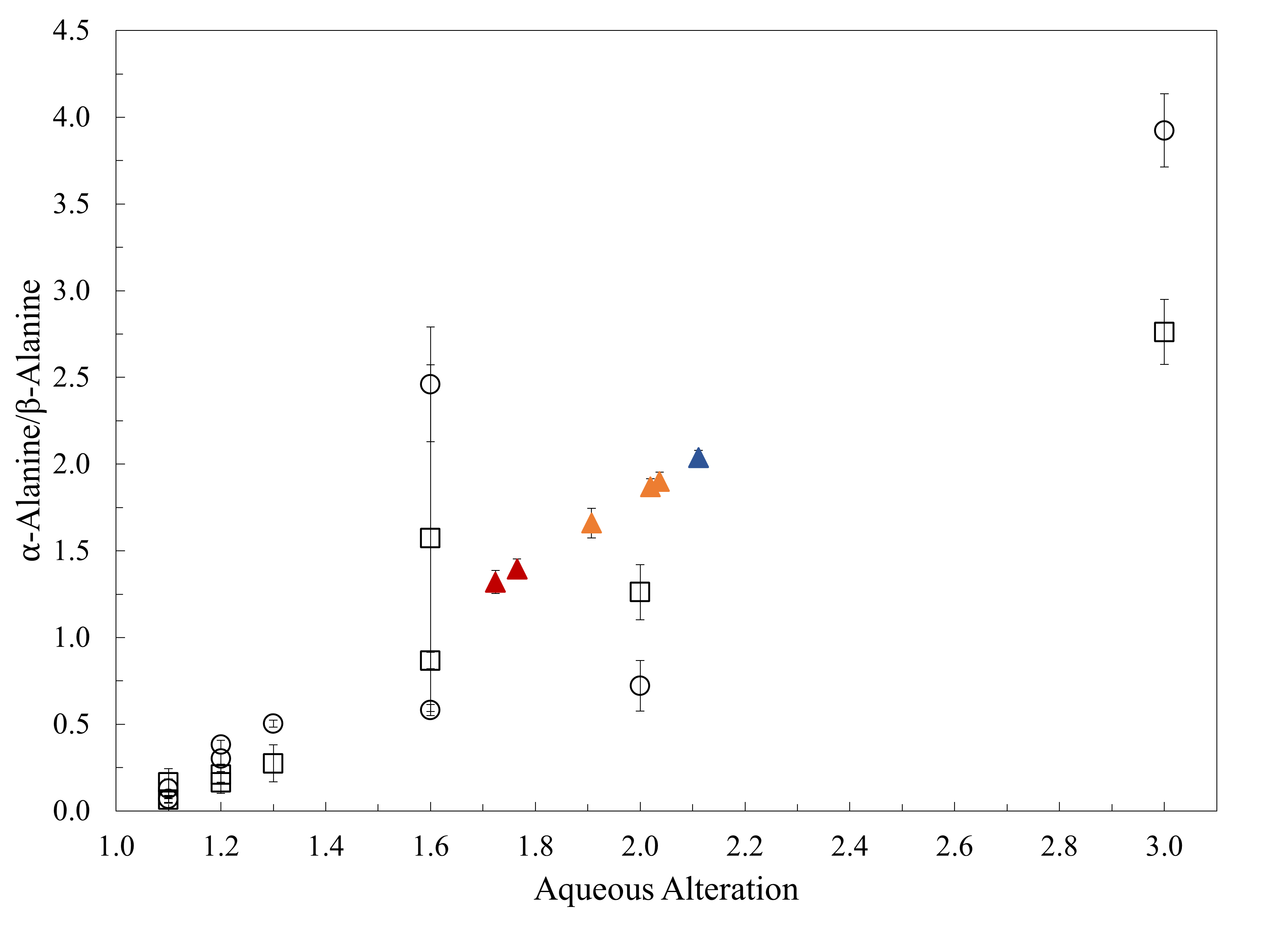}
\caption{The ratio of $\alpha$-alanine/$\beta$-alanine in unhydrolyzed (circles) and hydrolyzed (squares) extracts of various carbonaceous chondrites versus aqueous alteration,\citep{alexander2013classification} from unhydrolyzed and hydrolyzed amino acid data collected from the same hot water extract. The data from this study are plotted in the solid triangles along the empirical relationship between the alteration scale and the unhydrolyzed $\alpha$-alanine/$\beta$-alanine ratio, as described in the text. The initial $\alpha$-alanine/$\beta$-alanine in the three unheated residue samples is shown in the blue triangle. The orange and red triangles show the 50$^{\circ}$C 2, 7, and 30 days and the 125$^{\circ}$C 2 and 7 days heating experiments, respectively. Amino acid errors are propagated from the indicated sources. The meteorites plotted are hydrolyzed and unhydrolyzed CI1.1 Ivuna,\citep{burton2014effects} CI1.1 Orgueil,\citep{burton2014effects,glavin2010effects} CM1.2 SCO 06043,$^{\text{Supporting Information},}$\citep{burton2014effects} CM1.2 MET 01070,$^{\text{Supporting Information},}$\citep{burton2014effects} CR1.3 GRO 95577$^{\text{Supporting Information},}$\citep{burton2014effects} CM1.6 Murchison,\citep{glavin2020extraterrestrial} CM1.6 LEW 90500,\citep{glavin2006amino} C\textsubscript{ung}2.0 Tagish Lake (lithology 5b),\citep{glavin2014origin} and CM3.0 Asuka-12236.\citep{glavin2020abundant} Alexander et al.,\citep{alexander2013classification} for all meteorites with the exception of Tagish Lake 5b, which is described in Glavin et al.,\citep{glavin2020abundant} and Asuka-12236 from Nittler et al.\citep{nittler2020asuka}
}   
\label{fig2}
\end{figure}

Looking at the ratio of $\alpha$- to $\beta$-alanine (Table~\ref{table3}), there is a trend towards a lower ratio as the aqueous alteration experiment progresses in both time and temperature. Furthermore, taking the aqueous alteration scale from Alexander et al.\citep{alexander2013classification} for a variety of unhydrolyzed meteorites, there is an empirical relationship of $\alpha$-alanine/$\beta$-alanine = 1.85A-1.88, where A is the Alexander et al.\citep{alexander2013classification} alteration scale. A similar relationship for the Rubin et al.\citep{rubin2007progressive} scale can also be obtained (Supporting Information Figure S4). A comparison of these scales and their use in amino acid analyses are discussed by Glavin et al.\citep{glavin2020abundant} (the relationship for the hydrolyzed amino acids is similar) and were also used by Modica et al.\citep{modica2018amino} to compare the amino acid molecular distribution in laboratory-irradiated ice analogs with that of CM chondrites with different degrees of aqueous alteration. Figure~\ref{fig2} shows the relationship between $\alpha$-alanine/$\beta$-alanine and the aqueous alteration of various carbonaceous chondrites from the CI, CM and CR subgroups. The decrease in $\alpha$-alanine/$\beta$-alanine with time and, to a greater extent, temperature, follows the same trend observed in a range of increasingly aqueously altered carbonaceous chondrites. This appears to validate the applicability of the hydrothermal experiments.  

\begin{table*}
\centering
\begin{adjustbox}{width=1\textwidth}
	\caption{The ratio of $\alpha$-alanine/$\beta$-alanine from Table~\ref{table2}, with the errors propagated compared with the predicted aqueous alteration, which would produce the ratio as described in the text. The 30 days 125$^{\circ}$C ratio is not shown due to the pH drift and uncertain data quality as described in the text.}
	\label{table3}
	\begin{tabular}{c| c| c c c| c c c} 
		\hline
  \multicolumn{8}{c}{$\alpha$-alanine/$\beta$-alanine}\\
  \hline
       & Initial & 2 Days & 7 Days & 30 Days & 2 Days & 7 Days & 30 Days\\
& & & (50$^{\circ}$C)&  &&(125$^{\circ}$C) &\\ 
\hline
Residue A&2.02$\pm$0.02 & 1.87$\pm$0.05& -& - & 1.40$\pm$0.06 &- &-\\
\hline
Residue B&2.10$\pm$0.02 & -& 1.90$\pm$0.05& - & - &1.32$\pm$0.07 &-\\
\hline
Residue C&1.99$\pm$0.02 & -& -& 1.66$\pm$0.09 & - &- &-\\
\hline
Alteration Scale Equivalent & 2.1 & 2.0 & 2.0 & 1.9 & 1.8 & 1.7 &-\\
\hline

\end{tabular}
\end{adjustbox}
\end{table*}

\clearpage

The large presence of amines in the non-aqueously altered samples suggests that certain volatiles from the interstellar cloud can be efficiently transported to the protoplanetary disk, contributing to the inventory of aqueously altered interstellar-inherited species. In our experiments of the non-processed samples, amines from the ice experiments were trapped by and subsequently released from the residue, locked-up into salts during the ice experiments, and/or synthesized by the extraction step from potentially even more volatile species. Indeed, salts have been abundantly detected in comet 67P/Churyumov-Gerasimenko,\citep{altwegg2020evidence,altwegg2022abundant} in which their origin could be from irradiated interstellar ices. Therefore, in asteroids and comets, and potentially in other astrophysical bodies, it should be considered that in addition to refractory materials, certain volatiles from the ISM may also be present. 

\section{Conclusions}
\label{conclusions}

Experimental simulations of meteorite parent body aqueous alteration of interstellar residue analogs were performed to understand the impact of interstellar-inherited residues on the amine and amino acid distributions measured in meteorites. The results from this work suggest that the conditions in and organics from the parental interstellar cloud have an appreciable role in the abundances of meteoritic soluble organic matter, and that at least some of the interstellar organic abundances are likely altered in the meteorite parent body. This is supported by the following findings:   
\begin{itemize}

\item The abundance trend of methylamine $>$ ethylamine $>$ glycine $>$ serine $>$ alanine $>$ $\beta$-alanine remained constant before and after aqueous alteration, even with the presence of aqueous alteration reactants like cyanides, suggesting that the chemical inventory and conditions of the parental molecular cloud may be highly influential to the distribution of meteoritic amines and amino acids. Experiments with IOM-relevant materials, relevant minerals and inorganic species need to be performed to further evaluate the evolution of soluble organics from the interstellar cloud to the parent body.   

\item The abundances of ethylamine and alanine did not significantly change as a function of aqueous alteration, whereas the abundances of methylamine, glycine, serine, and $\beta$-alanine altered by $\sim$~twofold. The varying sensitives of each molecule to aqueous alteration may help explain the gap in linking interstellar to meteoritic abundances. The change in the ratio of $\alpha$-alanine/$\beta$-alanine with increasing aqueous alteration time and temperature was consistent with the trend observed in aqueously altered carbonaceous chondrites.  

\item Processing for 7 days at 125$^{\circ}$C resulted in the most growth of product abundances after aqueous alteration. 

\item The higher abundance of $\alpha$- than $\beta$-alanine matches well with results from aqueously processed commercial sample experiments, but is in contrast with what has been typically found in aqueously altered chondrites. Comparison of our results with that of other experimental simulations suggests that at least cyanide/nitriles reacting with relevant minerals may be key to linking interstellar to aqueously altered meteoritic amine and amino acid abundances. 

\item The relatively high abundances of volatile amines in the non-aqueously altered samples suggests that certain volatiles can be effectively delivered from the ISM to asteroids and comets, making a stronger case for the presence of interstellar organics beyond the ISM. 

\item Progress is needed from both, laboratory experiments and chondritic analyses, to better understand the link between interstellar and meteoritic amines and amino acids. 

\end{itemize}

\section{Supporting Information}
\label{SI}
More details on materials (S1), figures of amine and amino acid chromatograms (S2), tables of LC-FD/ToF-MS detection metrics and parameters (S3), table of unhydrolyzed meteorite amino acid data (S4), and figure of the ratio of $\alpha$-alanine/$\beta$-alanine as a function of aqueous alteration on the Rubin et al.\citep{rubin2007progressive} scale (S5) are provided.

\begin{acknowledgement}

This work was supported by the Emerging Worlds program, award number 19-EW19\textunderscore2-0021, by NASA under award number 80GSFC21M0002, the NASA Astrobiology Institute through award 13-13NAI7-0032 to the Goddard Center for Astrobiology (GCA), NASA’s Planetary Science Division Internal Scientist Funding Program through the Fundamental Laboratory Research (FLaRe) work package at NASA Goddard Space Flight Center, and the Simons Foundation (SCOL award 302497 to J.P.D.). D.Q. acknowledges Perry Gerakines for assistance with IR data analysis and Eric Parker for fruitful discussions. We acknowledge Stephen Brown, Eugene Gerashchenko, and Martin Carts for maintenance and operation of the Van de Graaff accelerator in the Radiation Effects Facility at GSFC.

\end{acknowledgement}

\bibliography{propyne.bib}

\providecommand{\latin}[1]{#1}
\makeatletter
\providecommand{\doi}
  {\begingroup\let\do\@makeother\dospecials
  \catcode`\{=1 \catcode`\}=2 \doi@aux}
\providecommand{\doi@aux}[1]{\endgroup\texttt{#1}}
\makeatother
\providecommand*\mcitethebibliography{\thebibliography}
\csname @ifundefined\endcsname{endmcitethebibliography}
  {\let\endmcitethebibliography\endthebibliography}{}
\begin{mcitethebibliography}{82}
\providecommand*\natexlab[1]{#1}
\providecommand*\mciteSetBstSublistMode[1]{}
\providecommand*\mciteSetBstMaxWidthForm[2]{}
\providecommand*\mciteBstWouldAddEndPuncttrue
  {\def\EndOfBibitem{\unskip.}}
\providecommand*\mciteBstWouldAddEndPunctfalse
  {\let\EndOfBibitem\relax}
\providecommand*\mciteSetBstMidEndSepPunct[3]{}
\providecommand*\mciteSetBstSublistLabelBeginEnd[3]{}
\providecommand*\EndOfBibitem{}
\mciteSetBstSublistMode{f}
\mciteSetBstMaxWidthForm{subitem}{(\alph{mcitesubitemcount})}
\mciteSetBstSublistLabelBeginEnd
  {\mcitemaxwidthsubitemform\space}
  {\relax}
  {\relax}

\bibitem[Glavin \latin{et~al.}(2018)Glavin, Alexander, Aponte, Dworkin, Elsila,
  and Yabuta]{glavin2018origin}
Glavin,~D.~P.; Alexander,~C.~M.; Aponte,~J.~C.; Dworkin,~J.~P.; Elsila,~J.~E.;
  Yabuta,~H. \emph{Primitive meteorites and asteroids}; Elsevier, 2018; pp
  205--271\relax
\mciteBstWouldAddEndPuncttrue
\mciteSetBstMidEndSepPunct{\mcitedefaultmidpunct}
{\mcitedefaultendpunct}{\mcitedefaultseppunct}\relax
\EndOfBibitem
\bibitem[Sandford \latin{et~al.}(2020)Sandford, Nuevo, Bera, and
  Lee]{sandford2020prebiotic}
Sandford,~S.~A.; Nuevo,~M.; Bera,~P.~P.; Lee,~T.~J. Prebiotic astrochemistry
  and the formation of molecules of astrobiological interest in interstellar
  clouds and protostellar disks. \emph{Chem. Rev.} \textbf{2020}, \emph{120},
  4616--4659\relax
\mciteBstWouldAddEndPuncttrue
\mciteSetBstMidEndSepPunct{\mcitedefaultmidpunct}
{\mcitedefaultendpunct}{\mcitedefaultseppunct}\relax
\EndOfBibitem
\bibitem[Materese \latin{et~al.}(2020)Materese, Gerakines, and
  Hudson]{materese2020laboratory}
Materese,~C.~K.; Gerakines,~P.~A.; Hudson,~R.~L. Laboratory Studies of
  Astronomical Ices: Reaction Chemistry and Spectroscopy. \emph{Acc. Chem.
  Res.} \textbf{2020}, \emph{54}, 280--290\relax
\mciteBstWouldAddEndPuncttrue
\mciteSetBstMidEndSepPunct{\mcitedefaultmidpunct}
{\mcitedefaultendpunct}{\mcitedefaultseppunct}\relax
\EndOfBibitem
\bibitem[Tachibana \latin{et~al.}(2014)Tachibana, Abe, Arakawa, Fujimoto,
  Iijima, Ishiguro, Kitazato, Kobayashi, Namiki, Okada, \latin{et~al.}
  others]{tachibana2014hayabusa2}
Tachibana,~S.; Abe,~M.; Arakawa,~M.; Fujimoto,~M.; Iijima,~Y.; Ishiguro,~M.;
  Kitazato,~K.; Kobayashi,~N.; Namiki,~N.; Okada,~T., \latin{et~al.}
  Hayabusa2: Scientific importance of samples returned from C-type near-Earth
  asteroid (162173) 1999 JU3. \emph{Geochem. J.} \textbf{2014}, \emph{48},
  571--587\relax
\mciteBstWouldAddEndPuncttrue
\mciteSetBstMidEndSepPunct{\mcitedefaultmidpunct}
{\mcitedefaultendpunct}{\mcitedefaultseppunct}\relax
\EndOfBibitem
\bibitem[Lauretta \latin{et~al.}(2017)Lauretta, Balram-Knutson, Beshore,
  Boynton, Drouet~d’Aubigny, DellaGiustina, Enos, Golish, Hergenrother,
  Howell, \latin{et~al.} others]{lauretta2017osiris}
Lauretta,~D.; Balram-Knutson,~S.; Beshore,~E.; Boynton,~W.;
  Drouet~d’Aubigny,~C.; DellaGiustina,~D.; Enos,~H.; Golish,~D.;
  Hergenrother,~C.; Howell,~E., \latin{et~al.}  OSIRIS-REx: sample return from
  asteroid (101955) Bennu. \emph{Space Sci. Rev.} \textbf{2017}, \emph{212},
  925--984\relax
\mciteBstWouldAddEndPuncttrue
\mciteSetBstMidEndSepPunct{\mcitedefaultmidpunct}
{\mcitedefaultendpunct}{\mcitedefaultseppunct}\relax
\EndOfBibitem
\bibitem[Martins \latin{et~al.}(2020)Martins, Chan, Bonal, King, and
  Yabuta]{martins2020organic}
Martins,~Z.; Chan,~Q. H.~S.; Bonal,~L.; King,~A.; Yabuta,~H. Organic matter in
  the solar system—implications for future on-site and sample return
  missions. \emph{Space Sci. Rev.} \textbf{2020}, \emph{216}, 1--23\relax
\mciteBstWouldAddEndPuncttrue
\mciteSetBstMidEndSepPunct{\mcitedefaultmidpunct}
{\mcitedefaultendpunct}{\mcitedefaultseppunct}\relax
\EndOfBibitem
\bibitem[Bernstein \latin{et~al.}(2002)Bernstein, Dworkin, Sandford, Cooper,
  and Allamandola]{bernstein2002racemic}
Bernstein,~M.~P.; Dworkin,~J.~P.; Sandford,~S.~A.; Cooper,~G.~W.;
  Allamandola,~L.~J. Racemic amino acids from the ultraviolet photolysis of
  interstellar ice analogues. \emph{Nature} \textbf{2002}, \emph{416},
  401--403\relax
\mciteBstWouldAddEndPuncttrue
\mciteSetBstMidEndSepPunct{\mcitedefaultmidpunct}
{\mcitedefaultendpunct}{\mcitedefaultseppunct}\relax
\EndOfBibitem
\bibitem[Munoz~Caro \latin{et~al.}(2002)Munoz~Caro, Meierhenrich, Schutte,
  Barbier, Arcones~Segovia, Rosenbauer, Thiemann, Brack, and
  Greenberg]{munoz2002amino}
Munoz~Caro,~G.; Meierhenrich,~U.~J.; Schutte,~W.~A.; Barbier,~B.;
  Arcones~Segovia,~A.; Rosenbauer,~H.; Thiemann,~W.-P.; Brack,~A.;
  Greenberg,~J.~M. Amino acids from ultraviolet irradiation of interstellar ice
  analogues. \emph{Nature} \textbf{2002}, \emph{416}, 403--406\relax
\mciteBstWouldAddEndPuncttrue
\mciteSetBstMidEndSepPunct{\mcitedefaultmidpunct}
{\mcitedefaultendpunct}{\mcitedefaultseppunct}\relax
\EndOfBibitem
\bibitem[Modica \latin{et~al.}(2018)Modica, Martins, Meinert, Zanda, and
  d’Hendecourt]{modica2018amino}
Modica,~P.; Martins,~Z.; Meinert,~C.; Zanda,~B.; d’Hendecourt,~L. The amino
  acid distribution in laboratory analogs of extraterrestrial organic matter: A
  comparison to CM chondrites. \emph{Astrophys. J.} \textbf{2018}, \emph{865},
  41\relax
\mciteBstWouldAddEndPuncttrue
\mciteSetBstMidEndSepPunct{\mcitedefaultmidpunct}
{\mcitedefaultendpunct}{\mcitedefaultseppunct}\relax
\EndOfBibitem
\bibitem[Martins \latin{et~al.}(2015)Martins, Modica, Zanda, and
  d'Hendecourt]{martins2015amino}
Martins,~Z.; Modica,~P.; Zanda,~B.; d'Hendecourt,~L. L.~S. The amino acid and
  hydrocarbon contents of the Paris meteorite: Insights into the most primitive
  CM chondrite. \emph{Meteorit. Planet. Sci.} \textbf{2015}, \emph{50},
  926--943\relax
\mciteBstWouldAddEndPuncttrue
\mciteSetBstMidEndSepPunct{\mcitedefaultmidpunct}
{\mcitedefaultendpunct}{\mcitedefaultseppunct}\relax
\EndOfBibitem
\bibitem[Elsila \latin{et~al.}(2007)Elsila, Dworkin, Bernstein, Martin, and
  Sandford]{elsila2007mechanisms}
Elsila,~J.~E.; Dworkin,~J.~P.; Bernstein,~M.~P.; Martin,~M.~P.; Sandford,~S.~A.
  Mechanisms of amino acid formation in interstellar ice analogs.
  \emph{Astrophys. J.} \textbf{2007}, \emph{660}, 911\relax
\mciteBstWouldAddEndPuncttrue
\mciteSetBstMidEndSepPunct{\mcitedefaultmidpunct}
{\mcitedefaultendpunct}{\mcitedefaultseppunct}\relax
\EndOfBibitem
\bibitem[Nuevo \latin{et~al.}(2008)Nuevo, Auger, Blanot, and
  d’Hendecourt]{nuevo2008detailed}
Nuevo,~M.; Auger,~G.; Blanot,~D.; d’Hendecourt,~L. A detailed study of the
  amino acids produced from the vacuum UV irradiation of interstellar ice
  analogs. \emph{Orig Life Evol Biosph.} \textbf{2008}, \emph{38}, 37--56\relax
\mciteBstWouldAddEndPuncttrue
\mciteSetBstMidEndSepPunct{\mcitedefaultmidpunct}
{\mcitedefaultendpunct}{\mcitedefaultseppunct}\relax
\EndOfBibitem
\bibitem[Aponte \latin{et~al.}(2014)Aponte, Dworkin, and
  Elsila]{aponte2014assessing}
Aponte,~J.~C.; Dworkin,~J.~P.; Elsila,~J.~E. Assessing the origins of aliphatic
  amines in the Murchison meteorite from their compound-specific carbon
  isotopic ratios and enantiomeric composition. \emph{Geochim. Cosmochim.
  Acta.} \textbf{2014}, \emph{141}, 331--345\relax
\mciteBstWouldAddEndPuncttrue
\mciteSetBstMidEndSepPunct{\mcitedefaultmidpunct}
{\mcitedefaultendpunct}{\mcitedefaultseppunct}\relax
\EndOfBibitem
\bibitem[Botta \latin{et~al.}(2002)Botta, Glavin, Kminek, and
  Bada]{botta2002relative}
Botta,~O.; Glavin,~D.~P.; Kminek,~G.; Bada,~J.~L. Relative amino acid
  concentrations as a signature for parent body processes of carbonaceous
  chondrites. \emph{Orig. Life Evol. Biosph} \textbf{2002}, \emph{32},
  143--163\relax
\mciteBstWouldAddEndPuncttrue
\mciteSetBstMidEndSepPunct{\mcitedefaultmidpunct}
{\mcitedefaultendpunct}{\mcitedefaultseppunct}\relax
\EndOfBibitem
\bibitem[Glavin \latin{et~al.}(2020)Glavin, McLain, Dworkin, Parker, Elsila,
  Aponte, Simkus, Pozarycki, Graham, Nittler, \latin{et~al.}
  others]{glavin2020abundant}
Glavin,~D.~P.; McLain,~H.~L.; Dworkin,~J.~P.; Parker,~E.~T.; Elsila,~J.~E.;
  Aponte,~J.~C.; Simkus,~D.~N.; Pozarycki,~C.~I.; Graham,~H.~V.;
  Nittler,~L.~R., \latin{et~al.}  Abundant extraterrestrial amino acids in the
  primitive CM carbonaceous chondrite Asuka 12236. \emph{Meteorit. Planet.
  Sci.} \textbf{2020}, \emph{55}, 1979--2006\relax
\mciteBstWouldAddEndPuncttrue
\mciteSetBstMidEndSepPunct{\mcitedefaultmidpunct}
{\mcitedefaultendpunct}{\mcitedefaultseppunct}\relax
\EndOfBibitem
\bibitem[Kebukawa \latin{et~al.}(2013)Kebukawa, Kilcoyne, and
  Cody]{kebukawa2013exploring}
Kebukawa,~Y.; Kilcoyne,~A.~D.; Cody,~G.~D. Exploring the potential formation of
  organic solids in chondrites and comets through polymerization of
  interstellar formaldehyde. \emph{Astrophys. J.} \textbf{2013}, \emph{771},
  19\relax
\mciteBstWouldAddEndPuncttrue
\mciteSetBstMidEndSepPunct{\mcitedefaultmidpunct}
{\mcitedefaultendpunct}{\mcitedefaultseppunct}\relax
\EndOfBibitem
\bibitem[Kebukawa \latin{et~al.}(2017)Kebukawa, Chan, Tachibana, Kobayashi, and
  Zolensky]{kebukawa2017one}
Kebukawa,~Y.; Chan,~Q.~H.; Tachibana,~S.; Kobayashi,~K.; Zolensky,~M.~E.
  One-pot synthesis of amino acid precursors with insoluble organic matter in
  planetesimals with aqueous activity. \emph{Sci. Adv.} \textbf{2017},
  \emph{3}, e1602093\relax
\mciteBstWouldAddEndPuncttrue
\mciteSetBstMidEndSepPunct{\mcitedefaultmidpunct}
{\mcitedefaultendpunct}{\mcitedefaultseppunct}\relax
\EndOfBibitem
\bibitem[Vinogradoff \latin{et~al.}(2018)Vinogradoff, Bernard, Le~Guillou, and
  Remusat]{vinogradoff2018evolution}
Vinogradoff,~V.; Bernard,~S.; Le~Guillou,~C.; Remusat,~L. Evolution of
  interstellar organic compounds under asteroidal hydrothermal conditions.
  \emph{Icarus} \textbf{2018}, \emph{305}, 358--370\relax
\mciteBstWouldAddEndPuncttrue
\mciteSetBstMidEndSepPunct{\mcitedefaultmidpunct}
{\mcitedefaultendpunct}{\mcitedefaultseppunct}\relax
\EndOfBibitem
\bibitem[Bernstein \latin{et~al.}(1995)Bernstein, Sandford, Allamandola, Chang,
  and Scharberg]{bernstein1995organic}
Bernstein,~M.~P.; Sandford,~S.~A.; Allamandola,~L.~J.; Chang,~S.;
  Scharberg,~M.~A. Organic compounds produced by photolysis of realistic
  interstellar and cometary ice analogs containing methanol. \emph{Astrophys.
  J.} \textbf{1995}, \emph{454}\relax
\mciteBstWouldAddEndPuncttrue
\mciteSetBstMidEndSepPunct{\mcitedefaultmidpunct}
{\mcitedefaultendpunct}{\mcitedefaultseppunct}\relax
\EndOfBibitem
\bibitem[Cottin \latin{et~al.}(2001)Cottin, Szopa, and
  Moore]{cottin2001production}
Cottin,~H.; Szopa,~C.; Moore,~M. Production of hexamethylenetetramine in
  photolyzed and irradiated interstellar cometary ice analogs. \emph{Astrophys.
  J.} \textbf{2001}, \emph{561}, L139\relax
\mciteBstWouldAddEndPuncttrue
\mciteSetBstMidEndSepPunct{\mcitedefaultmidpunct}
{\mcitedefaultendpunct}{\mcitedefaultseppunct}\relax
\EndOfBibitem
\bibitem[Caro and Schutte(2003)Caro, and Schutte]{caro2003uv}
Caro,~G.~M.; Schutte,~W. UV-photoprocessing of interstellar ice analogs: New
  infrared spectroscopic results. \emph{Astron. Astrophys.} \textbf{2003},
  \emph{412}, 121--132\relax
\mciteBstWouldAddEndPuncttrue
\mciteSetBstMidEndSepPunct{\mcitedefaultmidpunct}
{\mcitedefaultendpunct}{\mcitedefaultseppunct}\relax
\EndOfBibitem
\bibitem[Materese \latin{et~al.}(2020)Materese, Nuevo, Sandford, Bera, and
  Lee]{materese2020production}
Materese,~C.~K.; Nuevo,~M.; Sandford,~S.~A.; Bera,~P.~P.; Lee,~T.~J. The
  Production and Potential Detection of Hexamethylenetetramine-Methanol in
  Space. \emph{Astrobiology} \textbf{2020}, \emph{20}, 601--616\relax
\mciteBstWouldAddEndPuncttrue
\mciteSetBstMidEndSepPunct{\mcitedefaultmidpunct}
{\mcitedefaultendpunct}{\mcitedefaultseppunct}\relax
\EndOfBibitem
\bibitem[Danger \latin{et~al.}(2021)Danger, Vinogradoff, Matzka, Viennet,
  Remusat, Bernard, Ruf, Le~Sergeant~d’Hendecourt, and
  Schmitt-Kopplin]{danger2021exploring}
Danger,~G.; Vinogradoff,~V.; Matzka,~M.; Viennet,~J.; Remusat,~L.; Bernard,~S.;
  Ruf,~A.; Le~Sergeant~d’Hendecourt,~L.; Schmitt-Kopplin,~P. Exploring the
  link between molecular cloud ices and chondritic organic matter in
  laboratory. \emph{Nat. Commun.} \textbf{2021}, \emph{12}, 1--9\relax
\mciteBstWouldAddEndPuncttrue
\mciteSetBstMidEndSepPunct{\mcitedefaultmidpunct}
{\mcitedefaultendpunct}{\mcitedefaultseppunct}\relax
\EndOfBibitem
\bibitem[Hudson and Moore(2001)Hudson, and Moore]{hudson2001radiation}
Hudson,~R.; Moore,~M. Radiation chemical alterations in solar system ices: An
  overview. \emph{J. Geophys. Res. Planets} \textbf{2001}, \emph{106},
  33275--33284\relax
\mciteBstWouldAddEndPuncttrue
\mciteSetBstMidEndSepPunct{\mcitedefaultmidpunct}
{\mcitedefaultendpunct}{\mcitedefaultseppunct}\relax
\EndOfBibitem
\bibitem[Boogert \latin{et~al.}(2015)Boogert, Gerakines, and
  Whittet]{boogert2015observations}
Boogert,~A.~A.; Gerakines,~P.~A.; Whittet,~D.~C. Observations of the icy
  universe. \emph{Annu. Rev. Astron. Astrophys.} \textbf{2015}, \emph{53},
  541--581\relax
\mciteBstWouldAddEndPuncttrue
\mciteSetBstMidEndSepPunct{\mcitedefaultmidpunct}
{\mcitedefaultendpunct}{\mcitedefaultseppunct}\relax
\EndOfBibitem
\bibitem[Booth \latin{et~al.}(2021)Booth, Walsh, Terwisscha~van Scheltinga, van
  Dishoeck, Ilee, Hogerheijde, Kama, and Nomura]{booth2021inherited}
Booth,~A.~S.; Walsh,~C.; Terwisscha~van Scheltinga,~J.; van Dishoeck,~E.~F.;
  Ilee,~J.~D.; Hogerheijde,~M.~R.; Kama,~M.; Nomura,~H. An inherited complex
  organic molecule reservoir in a warm planet-hosting disk. \emph{Nat. Astron.}
  \textbf{2021}, \emph{5}, 684--690\relax
\mciteBstWouldAddEndPuncttrue
\mciteSetBstMidEndSepPunct{\mcitedefaultmidpunct}
{\mcitedefaultendpunct}{\mcitedefaultseppunct}\relax
\EndOfBibitem
\bibitem[Brunken \latin{et~al.}(2022)Brunken, Booth, Leemker, Nazari, van~der
  Marel, and van Dishoeck]{brunken2022major}
Brunken,~N.~G.; Booth,~A.~S.; Leemker,~M.; Nazari,~P.; van~der Marel,~N.; van
  Dishoeck,~E.~F. A major asymmetric ice trap in a planet-forming disk-III.
  First detection of dimethyl ether. \emph{Astron. Astrophys.} \textbf{2022},
  \emph{659}, A29\relax
\mciteBstWouldAddEndPuncttrue
\mciteSetBstMidEndSepPunct{\mcitedefaultmidpunct}
{\mcitedefaultendpunct}{\mcitedefaultseppunct}\relax
\EndOfBibitem
\bibitem[Zolensky \latin{et~al.}(1989)Zolensky, Bourcier, and
  Gooding]{zolensky1989aqueous}
Zolensky,~M.~E.; Bourcier,~W.~L.; Gooding,~J.~L. Aqueous alteration on the
  hydrous asteroids: Results of EQ3/6 computer simulations. \emph{Icarus}
  \textbf{1989}, \emph{78}, 411--425\relax
\mciteBstWouldAddEndPuncttrue
\mciteSetBstMidEndSepPunct{\mcitedefaultmidpunct}
{\mcitedefaultendpunct}{\mcitedefaultseppunct}\relax
\EndOfBibitem
\bibitem[Keil(2000)]{keil2000thermal}
Keil,~K. Thermal alteration of asteroids: evidence from meteorites.
  \emph{Planet. Space Sci.} \textbf{2000}, \emph{48}, 887--903\relax
\mciteBstWouldAddEndPuncttrue
\mciteSetBstMidEndSepPunct{\mcitedefaultmidpunct}
{\mcitedefaultendpunct}{\mcitedefaultseppunct}\relax
\EndOfBibitem
\bibitem[Guo and Eiler(2007)Guo, and Eiler]{guo2007temperatures}
Guo,~W.; Eiler,~J.~M. Temperatures of aqueous alteration and evidence for
  methane generation on the parent bodies of the CM chondrites. \emph{Geochim.
  Cosmochim. Acta.} \textbf{2007}, \emph{71}, 5565--5575\relax
\mciteBstWouldAddEndPuncttrue
\mciteSetBstMidEndSepPunct{\mcitedefaultmidpunct}
{\mcitedefaultendpunct}{\mcitedefaultseppunct}\relax
\EndOfBibitem
\bibitem[Weisberg \latin{et~al.}(2006)Weisberg, McCoy, Krot, \latin{et~al.}
  others]{weisberg2006systematics}
Weisberg,~M.~K.; McCoy,~T.~J.; Krot,~A.~N., \latin{et~al.}  Systematics and
  evaluation of meteorite classification. \emph{Meteorites and the early solar
  system II} \textbf{2006}, \emph{19}, 19--52\relax
\mciteBstWouldAddEndPuncttrue
\mciteSetBstMidEndSepPunct{\mcitedefaultmidpunct}
{\mcitedefaultendpunct}{\mcitedefaultseppunct}\relax
\EndOfBibitem
\bibitem[Brearley(2006)]{brearley2006action}
Brearley,~A.~J. The action of water. \emph{Meteorites and the early solar
  system II} \textbf{2006}, \emph{943}, 587--624\relax
\mciteBstWouldAddEndPuncttrue
\mciteSetBstMidEndSepPunct{\mcitedefaultmidpunct}
{\mcitedefaultendpunct}{\mcitedefaultseppunct}\relax
\EndOfBibitem
\bibitem[Elsila \latin{et~al.}(2016)Elsila, Aponte, Blackmond, Burton, Dworkin,
  and Glavin]{elsila2016meteoritic}
Elsila,~J.~E.; Aponte,~J.~C.; Blackmond,~D.~G.; Burton,~A.~S.; Dworkin,~J.~P.;
  Glavin,~D.~P. Meteoritic amino acids: Diversity in compositions reflects
  parent body histories. \emph{ACS Central Science} \textbf{2016}, \emph{2},
  370--379\relax
\mciteBstWouldAddEndPuncttrue
\mciteSetBstMidEndSepPunct{\mcitedefaultmidpunct}
{\mcitedefaultendpunct}{\mcitedefaultseppunct}\relax
\EndOfBibitem
\bibitem[Glavin \latin{et~al.}(2008)Glavin, Dworkin, and
  Sandford]{glavin2008detection}
Glavin,~D.~P.; Dworkin,~J.~P.; Sandford,~S.~A. Detection of cometary amines in
  samples returned by Stardust. \emph{Meteorit. Planet. Sci.} \textbf{2008},
  \emph{43}, 399--413\relax
\mciteBstWouldAddEndPuncttrue
\mciteSetBstMidEndSepPunct{\mcitedefaultmidpunct}
{\mcitedefaultendpunct}{\mcitedefaultseppunct}\relax
\EndOfBibitem
\bibitem[Burton \latin{et~al.}(2014)Burton, Grunsfeld, Elsila, Glavin, and
  Dworkin]{burton2014effects}
Burton,~A.~S.; Grunsfeld,~S.; Elsila,~J.~E.; Glavin,~D.~P.; Dworkin,~J.~P. The
  effects of parent-body hydrothermal heating on amino acid abundances in
  CI-like chondrites. \emph{Polar Sci.} \textbf{2014}, \emph{8}, 255--263\relax
\mciteBstWouldAddEndPuncttrue
\mciteSetBstMidEndSepPunct{\mcitedefaultmidpunct}
{\mcitedefaultendpunct}{\mcitedefaultseppunct}\relax
\EndOfBibitem
\bibitem[Aponte \latin{et~al.}(2020)Aponte, McLain, Simkus, Elsila, Glavin,
  Parker, Dworkin, Hill, Connolly~Jr, and Lauretta]{aponte2020extraterrestrial}
Aponte,~J.~C.; McLain,~H.~L.; Simkus,~D.~N.; Elsila,~J.~E.; Glavin,~D.~P.;
  Parker,~E.~T.; Dworkin,~J.~P.; Hill,~D.~H.; Connolly~Jr,~H.~C.;
  Lauretta,~D.~S. Extraterrestrial organic compounds and cyanide in the CM2
  carbonaceous chondrites Aguas Zarcas and Murchison. \emph{Meteorit. Planet.
  Sci.} \textbf{2020}, \emph{55}, 1509--1524\relax
\mciteBstWouldAddEndPuncttrue
\mciteSetBstMidEndSepPunct{\mcitedefaultmidpunct}
{\mcitedefaultendpunct}{\mcitedefaultseppunct}\relax
\EndOfBibitem
\bibitem[Altwegg \latin{et~al.}(2016)Altwegg, Balsiger, Bar-Nun, Berthelier,
  Bieler, Bochsler, Briois, Calmonte, Combi, Cottin, \latin{et~al.}
  others]{altwegg2016prebiotic}
Altwegg,~K.; Balsiger,~H.; Bar-Nun,~A.; Berthelier,~J.-J.; Bieler,~A.;
  Bochsler,~P.; Briois,~C.; Calmonte,~U.; Combi,~M.~R.; Cottin,~H.,
  \latin{et~al.}  Prebiotic chemicals—amino acid and phosphorus—in the coma
  of comet 67P/Churyumov-Gerasimenko. \emph{Sci. Adv.} \textbf{2016}, \emph{2},
  e1600285\relax
\mciteBstWouldAddEndPuncttrue
\mciteSetBstMidEndSepPunct{\mcitedefaultmidpunct}
{\mcitedefaultendpunct}{\mcitedefaultseppunct}\relax
\EndOfBibitem
\bibitem[Elsila \latin{et~al.}(2009)Elsila, Glavin, and
  Dworkin]{elsila2009cometary}
Elsila,~J.~E.; Glavin,~D.~P.; Dworkin,~J.~P. Cometary glycine detected in
  samples returned by Stardust. \emph{Meteorit. Planet. Sci.} \textbf{2009},
  \emph{44}, 1323--1330\relax
\mciteBstWouldAddEndPuncttrue
\mciteSetBstMidEndSepPunct{\mcitedefaultmidpunct}
{\mcitedefaultendpunct}{\mcitedefaultseppunct}\relax
\EndOfBibitem
\bibitem[Kaifu \latin{et~al.}(1974)Kaifu, Morimoto, Nagane, Akabane, Iguchi,
  and Takagi]{kaifu1974detection}
Kaifu,~N.; Morimoto,~M.; Nagane,~K.; Akabane,~K.; Iguchi,~T.; Takagi,~K.
  Detection of interstellar methylamine. \emph{Astrophys. J.} \textbf{1974},
  \emph{191}, L135--L137\relax
\mciteBstWouldAddEndPuncttrue
\mciteSetBstMidEndSepPunct{\mcitedefaultmidpunct}
{\mcitedefaultendpunct}{\mcitedefaultseppunct}\relax
\EndOfBibitem
\bibitem[B{\o}gelund \latin{et~al.}(2019)B{\o}gelund, McGuire, Hogerheijde, van
  Dishoeck, and Ligterink]{bogelund2019methylamine}
B{\o}gelund,~E.~G.; McGuire,~B.~A.; Hogerheijde,~M.~R.; van Dishoeck,~E.~F.;
  Ligterink,~N.~F. Methylamine and other simple N-bearing species in the hot
  cores NGC 6334I MM1--3. \emph{Astron. Astrophys.} \textbf{2019}, \emph{624},
  A82\relax
\mciteBstWouldAddEndPuncttrue
\mciteSetBstMidEndSepPunct{\mcitedefaultmidpunct}
{\mcitedefaultendpunct}{\mcitedefaultseppunct}\relax
\EndOfBibitem
\bibitem[Ehrenfreund \latin{et~al.}(2001)Ehrenfreund, Glavin, Botta, Cooper,
  and Bada]{ehrenfreund2001extraterrestrial}
Ehrenfreund,~P.; Glavin,~D.~P.; Botta,~O.; Cooper,~G.; Bada,~J.~L.
  Extraterrestrial amino acids in Orgueil and Ivuna: Tracing the parent body of
  CI type carbonaceous chondrites. \emph{Proc. Natl. Acad. Sci. U.S.A.}
  \textbf{2001}, \emph{98}, 2138--2141\relax
\mciteBstWouldAddEndPuncttrue
\mciteSetBstMidEndSepPunct{\mcitedefaultmidpunct}
{\mcitedefaultendpunct}{\mcitedefaultseppunct}\relax
\EndOfBibitem
\bibitem[Hudson and Moore(1999)Hudson, and Moore]{hudson1999laboratory}
Hudson,~R.; Moore,~M. Laboratory studies of the formation of methanol and other
  organic molecules by water + carbon monoxide radiolysis: Relevance to comets,
  icy satellites, and interstellar ices. \emph{Icarus} \textbf{1999},
  \emph{140}, 451--461\relax
\mciteBstWouldAddEndPuncttrue
\mciteSetBstMidEndSepPunct{\mcitedefaultmidpunct}
{\mcitedefaultendpunct}{\mcitedefaultseppunct}\relax
\EndOfBibitem
\bibitem[Gerakines and Hudson(2013)Gerakines, and Hudson]{gerakines2013glycine}
Gerakines,~P.~A.; Hudson,~R.~L. Glycine's radiolytic destruction in ices: first
  in situ laboratory measurements for Mars. \emph{Astrobiology} \textbf{2013},
  \emph{13}, 647--655\relax
\mciteBstWouldAddEndPuncttrue
\mciteSetBstMidEndSepPunct{\mcitedefaultmidpunct}
{\mcitedefaultendpunct}{\mcitedefaultseppunct}\relax
\EndOfBibitem
\bibitem[Gerakines \latin{et~al.}(2022)Gerakines, Qasim, Frail, and
  Hudson]{gerakines2022radiolytic}
Gerakines,~P.~A.; Qasim,~D.; Frail,~S.; Hudson,~R.~L. Radiolytic Destruction of
  Uracil in Interstellar and Solar System Ices. \emph{Astrobiology}
  \textbf{2022}, \emph{22}, 233--241\relax
\mciteBstWouldAddEndPuncttrue
\mciteSetBstMidEndSepPunct{\mcitedefaultmidpunct}
{\mcitedefaultendpunct}{\mcitedefaultseppunct}\relax
\EndOfBibitem
\bibitem[Heavens(2011)]{heavens1991optical}
Heavens,~O.~S. \emph{Optical Properties of Thin Solid Films}, 2nd ed.; 2011; p
  114\relax
\mciteBstWouldAddEndPuncttrue
\mciteSetBstMidEndSepPunct{\mcitedefaultmidpunct}
{\mcitedefaultendpunct}{\mcitedefaultseppunct}\relax
\EndOfBibitem
\bibitem[Weast \latin{et~al.}(1984)Weast, Suby, and Hodman]{weast1984handbook}
Weast,~R.~C.; Suby,~S.; Hodman,~C. \emph{Handbook of Chemistry and Physics.
  64}; 1984\relax
\mciteBstWouldAddEndPuncttrue
\mciteSetBstMidEndSepPunct{\mcitedefaultmidpunct}
{\mcitedefaultendpunct}{\mcitedefaultseppunct}\relax
\EndOfBibitem
\bibitem[Narten \latin{et~al.}(1976)Narten, Venkatesh, and
  Rice]{narten1976diffraction}
Narten,~A.; Venkatesh,~C.-G.; Rice,~S. Diffraction pattern and structure of
  amorphous solid water at 10 and 77 K. \emph{J. Chem. Phys.} \textbf{1976},
  \emph{64}, 1106\relax
\mciteBstWouldAddEndPuncttrue
\mciteSetBstMidEndSepPunct{\mcitedefaultmidpunct}
{\mcitedefaultendpunct}{\mcitedefaultseppunct}\relax
\EndOfBibitem
\bibitem[Loeffler \latin{et~al.}(2016)Loeffler, Moore, and
  Gerakines]{loeffler2016effects}
Loeffler,~M.; Moore,~M.; Gerakines,~P. The effects of experimental conditions
  on the refractive index and density of low-temperature ices: solid carbon
  dioxide. \emph{Astrophys. J.} \textbf{2016}, \emph{827}, 98\relax
\mciteBstWouldAddEndPuncttrue
\mciteSetBstMidEndSepPunct{\mcitedefaultmidpunct}
{\mcitedefaultendpunct}{\mcitedefaultseppunct}\relax
\EndOfBibitem
\bibitem[Luna \latin{et~al.}(2018)Luna, Molpeceres, Ortigoso, Satorre, Domingo,
  and Mat{\'e}]{luna2018densities}
Luna,~R.; Molpeceres,~G.; Ortigoso,~J.; Satorre,~M.~A.; Domingo,~M.;
  Mat{\'e},~B. Densities, infrared band strengths, and optical constants of
  solid methanol. \emph{Astron. Astrophys.} \textbf{2018}, \emph{617},
  A116\relax
\mciteBstWouldAddEndPuncttrue
\mciteSetBstMidEndSepPunct{\mcitedefaultmidpunct}
{\mcitedefaultendpunct}{\mcitedefaultseppunct}\relax
\EndOfBibitem
\bibitem[Bouilloud \latin{et~al.}(2015)Bouilloud, Fray, B{\'e}nilan, Cottin,
  Gazeau, and Jolly]{bouilloud2015bibliographic}
Bouilloud,~M.; Fray,~N.; B{\'e}nilan,~Y.; Cottin,~H.; Gazeau,~M.-C.; Jolly,~A.
  Bibliographic review and new measurements of the infrared band strengths of
  pure molecules at 25 K: H$_2$O, CO$_2$, CO, CH$_4$, NH$_3$, CH$_3$OH, HCOOH
  and H$_2$CO. \emph{Mon. Not. R. Astron. Soc.} \textbf{2015}, \emph{451},
  2145--2160\relax
\mciteBstWouldAddEndPuncttrue
\mciteSetBstMidEndSepPunct{\mcitedefaultmidpunct}
{\mcitedefaultendpunct}{\mcitedefaultseppunct}\relax
\EndOfBibitem
\bibitem[Moore \latin{et~al.}(2001)Moore, Hudson, and Gerakines]{moore2001mid}
Moore,~M.; Hudson,~R.; Gerakines,~P. Mid-and far-infrared spectroscopic studies
  of the influence of temperature, ultraviolet photolysis and ion irradiation
  on cosmic-type ices. \emph{Spectrochim. Acta A Mol. Biomol. Spectrosc.}
  \textbf{2001}, \emph{57}, 843--858\relax
\mciteBstWouldAddEndPuncttrue
\mciteSetBstMidEndSepPunct{\mcitedefaultmidpunct}
{\mcitedefaultendpunct}{\mcitedefaultseppunct}\relax
\EndOfBibitem
\bibitem[Ziegler \latin{et~al.}(2010)Ziegler, Ziegler, and
  Biersack]{ziegler2010srim}
Ziegler,~J.~F.; Ziegler,~M.~D.; Biersack,~J.~P. SRIM--The stopping and range of
  ions in matter (2010). \emph{Nucl. Instrum. Methods. Phys. Res. B}
  \textbf{2010}, \emph{268}, 1818--1823\relax
\mciteBstWouldAddEndPuncttrue
\mciteSetBstMidEndSepPunct{\mcitedefaultmidpunct}
{\mcitedefaultendpunct}{\mcitedefaultseppunct}\relax
\EndOfBibitem
\bibitem[Dworkin \latin{et~al.}(2018)Dworkin, Adelman, Ajluni, Andronikov,
  Aponte, Bartels, Beshore, Bierhaus, Brucato, Bryan, \latin{et~al.}
  others]{dworkin2018osiris}
Dworkin,~J.; Adelman,~L.; Ajluni,~T.; Andronikov,~A.; Aponte,~J.; Bartels,~A.;
  Beshore,~E.; Bierhaus,~E.; Brucato,~J.; Bryan,~B., \latin{et~al.}  OSIRIS-REx
  contamination control strategy and implementation. \emph{Space Sci. Rev.}
  \textbf{2018}, \emph{214}, 1--53\relax
\mciteBstWouldAddEndPuncttrue
\mciteSetBstMidEndSepPunct{\mcitedefaultmidpunct}
{\mcitedefaultendpunct}{\mcitedefaultseppunct}\relax
\EndOfBibitem
\bibitem[Pavlov \latin{et~al.}(2022)Pavlov, McLain, Glavin, Roussel, Dworkin,
  Elsila, and Yocum]{pavlov2022rapid}
Pavlov,~A.~A.; McLain,~H.~L.; Glavin,~D.~P.; Roussel,~A.; Dworkin,~J.~P.;
  Elsila,~J.~E.; Yocum,~K.~M. Rapid Radiolytic Degradation of Amino Acids in
  the Martian Shallow Subsurface: Implications for the Search for Extinct Life.
  \emph{Astrobiology} \textbf{2022}, \relax
\mciteBstWouldAddEndPunctfalse
\mciteSetBstMidEndSepPunct{\mcitedefaultmidpunct}
{}{\mcitedefaultseppunct}\relax
\EndOfBibitem
\bibitem[Simkus \latin{et~al.}(2019)Simkus, Aponte, Elsila, Parker, Glavin, and
  Dworkin]{simkus2019methodologies}
Simkus,~D.~N.; Aponte,~J.~C.; Elsila,~J.~E.; Parker,~E.~T.; Glavin,~D.~P.;
  Dworkin,~J.~P. Methodologies for analyzing soluble organic compounds in
  extraterrestrial samples: Amino acids, amines, monocarboxylic acids,
  aldehydes, and ketones. \emph{Life} \textbf{2019}, \emph{9}, 47\relax
\mciteBstWouldAddEndPuncttrue
\mciteSetBstMidEndSepPunct{\mcitedefaultmidpunct}
{\mcitedefaultendpunct}{\mcitedefaultseppunct}\relax
\EndOfBibitem
\bibitem[F{\"o}rstel \latin{et~al.}(2017)F{\"o}rstel, Bergantini, Maksyutenko,
  G{\'o}bi, and Kaiser]{forstel2017formation}
F{\"o}rstel,~M.; Bergantini,~A.; Maksyutenko,~P.; G{\'o}bi,~S.; Kaiser,~R.~I.
  Formation of methylamine and ethylamine in extraterrestrial ices and their
  role as fundamental building blocks of proteinogenic $\alpha$-amino acids.
  \emph{Astrophys. J.} \textbf{2017}, \emph{845}, 83\relax
\mciteBstWouldAddEndPuncttrue
\mciteSetBstMidEndSepPunct{\mcitedefaultmidpunct}
{\mcitedefaultendpunct}{\mcitedefaultseppunct}\relax
\EndOfBibitem
\bibitem[Moore and Khanna(1991)Moore, and Khanna]{moore1991infrared}
Moore,~M.; Khanna,~R. Infrared and mass spectral studies of proton irradiated
  H$_2$O + CO$_2$ ice: evidence for carbonic acid. \emph{Spectrochim. Acta A
  Mol. Spectrosc.} \textbf{1991}, \emph{47}, 255--262\relax
\mciteBstWouldAddEndPuncttrue
\mciteSetBstMidEndSepPunct{\mcitedefaultmidpunct}
{\mcitedefaultendpunct}{\mcitedefaultseppunct}\relax
\EndOfBibitem
\bibitem[Palumbo \latin{et~al.}(2000)Palumbo, Strazzulla, Pendleton, and
  Tielens]{palumbo2000roc}
Palumbo,~M.; Strazzulla,~G.; Pendleton,~Y.; Tielens,~A. ROC$\equiv$N Species
  Produced by Ion Irradiation of Ice Mixtures: Comparison with Astronomical
  Observations. \emph{Astrophys. J.} \textbf{2000}, \emph{534}, 801\relax
\mciteBstWouldAddEndPuncttrue
\mciteSetBstMidEndSepPunct{\mcitedefaultmidpunct}
{\mcitedefaultendpunct}{\mcitedefaultseppunct}\relax
\EndOfBibitem
\bibitem[Danger \latin{et~al.}(2011)Danger, Bossa, De~Marcellus, Borget,
  Duvernay, Theul{\'e}, Chiavassa, and d’Hendecourt]{danger2011experimental}
Danger,~G.; Bossa,~J.-B.; De~Marcellus,~P.; Borget,~F.; Duvernay,~F.;
  Theul{\'e},~P.; Chiavassa,~T.; d’Hendecourt,~L. Experimental investigation
  of nitrile formation from VUV photochemistry of interstellar ices analogs:
  acetonitrile and amino acetonitrile. \emph{Astron. Astrophys.} \textbf{2011},
  \emph{525}, A30\relax
\mciteBstWouldAddEndPuncttrue
\mciteSetBstMidEndSepPunct{\mcitedefaultmidpunct}
{\mcitedefaultendpunct}{\mcitedefaultseppunct}\relax
\EndOfBibitem
\bibitem[Pardo \latin{et~al.}(1993)Pardo, Osman, Weinstein, and
  Rabinowitz]{pardo1993mechanisms}
Pardo,~L.; Osman,~R.; Weinstein,~H.; Rabinowitz,~J.~R. Mechanisms of
  nucleophilic addition to activated double bonds: 1, 2- and 1, 4-Michael
  addition of ammonia. \emph{J. Am. Chem. Soc.} \textbf{1993}, \emph{115},
  8263--8269\relax
\mciteBstWouldAddEndPuncttrue
\mciteSetBstMidEndSepPunct{\mcitedefaultmidpunct}
{\mcitedefaultendpunct}{\mcitedefaultseppunct}\relax
\EndOfBibitem
\bibitem[Magrino \latin{et~al.}(2021)Magrino, Pietrucci, and
  Saitta]{magrino2021step}
Magrino,~T.; Pietrucci,~F.; Saitta,~A.~M. Step by step strecker amino acid
  synthesis from ab initio prebiotic chemistry. \emph{J. Phys. Chem. Lett.}
  \textbf{2021}, \emph{12}, 2630--2637\relax
\mciteBstWouldAddEndPuncttrue
\mciteSetBstMidEndSepPunct{\mcitedefaultmidpunct}
{\mcitedefaultendpunct}{\mcitedefaultseppunct}\relax
\EndOfBibitem
\bibitem[Briggs \latin{et~al.}(1992)Briggs, Ertem, Ferris, Greenberg, McCain,
  Mendoza-Gomez, and Schutte]{briggs1992comet}
Briggs,~R.; Ertem,~G.; Ferris,~J.; Greenberg,~J.; McCain,~P.;
  Mendoza-Gomez,~C.; Schutte,~W. Comet Halley as an aggregate of interstellar
  dust and further evidence for the photochemical formation of organics in the
  interstellar medium. \emph{Orig. Life Evol. Biosph.} \textbf{1992},
  \emph{22}, 287--307\relax
\mciteBstWouldAddEndPuncttrue
\mciteSetBstMidEndSepPunct{\mcitedefaultmidpunct}
{\mcitedefaultendpunct}{\mcitedefaultseppunct}\relax
\EndOfBibitem
\bibitem[Vinogradoff \latin{et~al.}(2013)Vinogradoff, Fray, Duvernay, Briani,
  Danger, Cottin, Theul{\'e}, and Chiavassa]{vinogradoff2013importance}
Vinogradoff,~V.; Fray,~N.; Duvernay,~F.; Briani,~G.; Danger,~G.; Cottin,~H.;
  Theul{\'e},~P.; Chiavassa,~T. Importance of thermal reactivity for
  hexamethylenetetramine formation from simulated interstellar ices.
  \emph{Astron. Astrophys.} \textbf{2013}, \emph{551}, A128\relax
\mciteBstWouldAddEndPuncttrue
\mciteSetBstMidEndSepPunct{\mcitedefaultmidpunct}
{\mcitedefaultendpunct}{\mcitedefaultseppunct}\relax
\EndOfBibitem
\bibitem[Weiss \latin{et~al.}(2018)Weiss, Muth, Drumm, and
  Kirchner]{weiss2018thermal}
Weiss,~I.~M.; Muth,~C.; Drumm,~R.; Kirchner,~H.~O. Thermal decomposition of the
  amino acids glycine, cysteine, aspartic acid, asparagine, glutamic acid,
  glutamine, arginine and histidine. \emph{BMC Biophys.} \textbf{2018},
  \emph{11}, 1--15\relax
\mciteBstWouldAddEndPuncttrue
\mciteSetBstMidEndSepPunct{\mcitedefaultmidpunct}
{\mcitedefaultendpunct}{\mcitedefaultseppunct}\relax
\EndOfBibitem
\bibitem[Rodante(1992)]{rodante1992thermodynamics}
Rodante,~F. Thermodynamics and kinetics of decomposition processes for standard
  $\alpha$-amino acids and some of their dipeptides in the solid state.
  \emph{Thermochim. Acta} \textbf{1992}, \emph{200}, 47--61\relax
\mciteBstWouldAddEndPuncttrue
\mciteSetBstMidEndSepPunct{\mcitedefaultmidpunct}
{\mcitedefaultendpunct}{\mcitedefaultseppunct}\relax
\EndOfBibitem
\bibitem[Vinogradoff \latin{et~al.}(2020)Vinogradoff, Remusat, McLain, Aponte,
  Bernard, Danger, Dworkin, Elsila, and Jaber]{vinogradoff2020impact}
Vinogradoff,~V.; Remusat,~L.; McLain,~H.; Aponte,~J.; Bernard,~S.; Danger,~G.;
  Dworkin,~J.; Elsila,~J.; Jaber,~M. Impact of phyllosilicates on amino acid
  formation under asteroidal conditions. \emph{ACS Earth Space Chem.}
  \textbf{2020}, \emph{4}, 1398--1407\relax
\mciteBstWouldAddEndPuncttrue
\mciteSetBstMidEndSepPunct{\mcitedefaultmidpunct}
{\mcitedefaultendpunct}{\mcitedefaultseppunct}\relax
\EndOfBibitem
\bibitem[Breslow(1959)]{breslow1959mechanism}
Breslow,~R. On the mechanism of the formose reaction. \emph{Tetrahedron Lett.}
  \textbf{1959}, \emph{1}, 22--26\relax
\mciteBstWouldAddEndPuncttrue
\mciteSetBstMidEndSepPunct{\mcitedefaultmidpunct}
{\mcitedefaultendpunct}{\mcitedefaultseppunct}\relax
\EndOfBibitem
\bibitem[Kopetzki and Antonietti(2011)Kopetzki, and
  Antonietti]{kopetzki2011hydrothermal}
Kopetzki,~D.; Antonietti,~M. Hydrothermal formose reaction. \emph{New J. Chem.}
  \textbf{2011}, \emph{35}, 1787--1794\relax
\mciteBstWouldAddEndPuncttrue
\mciteSetBstMidEndSepPunct{\mcitedefaultmidpunct}
{\mcitedefaultendpunct}{\mcitedefaultseppunct}\relax
\EndOfBibitem
\bibitem[Hudson \latin{et~al.}(2005)Hudson, Moore, and Cook]{hudson2005ir}
Hudson,~R.~L.; Moore,~M.~H.; Cook,~A.~M. IR characterization and radiation
  chemistry of glycolaldehyde and ethylene glycol ices. \emph{Adv. Space Res.}
  \textbf{2005}, \emph{36}, 184--189\relax
\mciteBstWouldAddEndPuncttrue
\mciteSetBstMidEndSepPunct{\mcitedefaultmidpunct}
{\mcitedefaultendpunct}{\mcitedefaultseppunct}\relax
\EndOfBibitem
\bibitem[de~Marcellus \latin{et~al.}(2015)de~Marcellus, Meinert, Myrgorodska,
  Nahon, Buhse, d’Hendecourt, and Meierhenrich]{de2015aldehydes}
de~Marcellus,~P.; Meinert,~C.; Myrgorodska,~I.; Nahon,~L.; Buhse,~T.;
  d’Hendecourt,~L. L.~S.; Meierhenrich,~U.~J. Aldehydes and sugars from
  evolved precometary ice analogs: Importance of ices in astrochemical and
  prebiotic evolution. \emph{Proc. Natl. Acad. Sci. U.S.A.} \textbf{2015},
  \emph{112}, 965--970\relax
\mciteBstWouldAddEndPuncttrue
\mciteSetBstMidEndSepPunct{\mcitedefaultmidpunct}
{\mcitedefaultendpunct}{\mcitedefaultseppunct}\relax
\EndOfBibitem
\bibitem[Fedoseev \latin{et~al.}(2015)Fedoseev, Cuppen, Ioppolo, Lamberts, and
  Linnartz]{fedoseev2015experimental}
Fedoseev,~G.; Cuppen,~H.~M.; Ioppolo,~S.; Lamberts,~T.; Linnartz,~H.
  Experimental evidence for glycolaldehyde and ethylene glycol formation by
  surface hydrogenation of CO molecules under dense molecular cloud conditions.
  \emph{Mon. Not. R. Astron. Soc.} \textbf{2015}, \emph{448}, 1288--1297\relax
\mciteBstWouldAddEndPuncttrue
\mciteSetBstMidEndSepPunct{\mcitedefaultmidpunct}
{\mcitedefaultendpunct}{\mcitedefaultseppunct}\relax
\EndOfBibitem
\bibitem[Parker \latin{et~al.}(2022)Parker, Furusho, Glavin, Hamase, Naraoka,
  Takano, and Dworkin]{parker2022amino}
Parker,~E.; Furusho,~A.; Glavin,~D.; Hamase,~K.; Naraoka,~H.; Takano,~Y.;
  Dworkin,~J. Amino Acid Analyses of a Sample of Ryugu by a Combination of
  Liquid Chromatography and High-Resolution Mass Spectrometry Techniques.
  \emph{LPI Contributions} \textbf{2022}, \emph{2678}, 2651\relax
\mciteBstWouldAddEndPuncttrue
\mciteSetBstMidEndSepPunct{\mcitedefaultmidpunct}
{\mcitedefaultendpunct}{\mcitedefaultseppunct}\relax
\EndOfBibitem
\bibitem[Alexander \latin{et~al.}(2013)Alexander, Howard, Bowden, and
  Fogel]{alexander2013classification}
Alexander,~C.~M.; Howard,~K.~T.; Bowden,~R.; Fogel,~M.~L. The classification of
  CM and CR chondrites using bulk H, C and N abundances and isotopic
  compositions. \emph{Geochim. Cosmochim. Acta.} \textbf{2013}, \emph{123},
  244--260\relax
\mciteBstWouldAddEndPuncttrue
\mciteSetBstMidEndSepPunct{\mcitedefaultmidpunct}
{\mcitedefaultendpunct}{\mcitedefaultseppunct}\relax
\EndOfBibitem
\bibitem[Rubin \latin{et~al.}(2007)Rubin, Trigo-Rodr{\'\i}guez, Huber, and
  Wasson]{rubin2007progressive}
Rubin,~A.~E.; Trigo-Rodr{\'\i}guez,~J.~M.; Huber,~H.; Wasson,~J.~T. Progressive
  aqueous alteration of CM carbonaceous chondrites. \emph{Geochim. Cosmochim.
  Acta.} \textbf{2007}, \emph{71}, 2361--2382\relax
\mciteBstWouldAddEndPuncttrue
\mciteSetBstMidEndSepPunct{\mcitedefaultmidpunct}
{\mcitedefaultendpunct}{\mcitedefaultseppunct}\relax
\EndOfBibitem
\bibitem[Glavin \latin{et~al.}(2010)Glavin, Callahan, Dworkin, and
  Elsila]{glavin2010effects}
Glavin,~D.~P.; Callahan,~M.~P.; Dworkin,~J.~P.; Elsila,~J.~E. The effects of
  parent body processes on amino acids in carbonaceous chondrites.
  \emph{Meteorit. Planet. Sci.} \textbf{2010}, \emph{45}, 1948--1972\relax
\mciteBstWouldAddEndPuncttrue
\mciteSetBstMidEndSepPunct{\mcitedefaultmidpunct}
{\mcitedefaultendpunct}{\mcitedefaultseppunct}\relax
\EndOfBibitem
\bibitem[Glavin \latin{et~al.}(2020)Glavin, Elsila, McLain, Aponte, Parker,
  Dworkin, Hill, Connolly~Jr, and Lauretta]{glavin2020extraterrestrial}
Glavin,~D.~P.; Elsila,~J.~E.; McLain,~H.~L.; Aponte,~J.~C.; Parker,~E.~T.;
  Dworkin,~J.~P.; Hill,~D.~H.; Connolly~Jr,~H.~C.; Lauretta,~D.~S.
  Extraterrestrial amino acids and L-enantiomeric excesses in the CM 2
  carbonaceous chondrites Aguas Zarcas and Murchison. \emph{Meteorit. Planet.
  Sci.} \textbf{2020}, \emph{56}, 148--173\relax
\mciteBstWouldAddEndPuncttrue
\mciteSetBstMidEndSepPunct{\mcitedefaultmidpunct}
{\mcitedefaultendpunct}{\mcitedefaultseppunct}\relax
\EndOfBibitem
\bibitem[Glavin \latin{et~al.}(2006)Glavin, Dworkin, Aubrey, Botta, Doty~III,
  Martins, and Bada]{glavin2006amino}
Glavin,~D.~P.; Dworkin,~J.~P.; Aubrey,~A.; Botta,~O.; Doty~III,~J.~H.;
  Martins,~Z.; Bada,~J.~L. Amino acid analyses of Antarctic CM2 meteorites
  using liquid chromatography-time of flight-mass spectrometry. \emph{Meteorit.
  Planet. Sci.} \textbf{2006}, \emph{41}, 889--902\relax
\mciteBstWouldAddEndPuncttrue
\mciteSetBstMidEndSepPunct{\mcitedefaultmidpunct}
{\mcitedefaultendpunct}{\mcitedefaultseppunct}\relax
\EndOfBibitem
\bibitem[Glavin \latin{et~al.}(2014)Glavin, Freissinet, Eigenbrode, Miller,
  Martin, Summons, Steele, Franz, Archer, Brinckerhoff, \latin{et~al.}
  others]{glavin2014origin}
Glavin,~D.; Freissinet,~C.; Eigenbrode,~J.; Miller,~K.; Martin,~M.;
  Summons,~R.; Steele,~A.; Franz,~H.; Archer,~D.; Brinckerhoff,~W.,
  \latin{et~al.}  Origin of Chlorobenzene Detected by the Curiosity Rover in
  Yellowknife Bay: Evidence for Martian Organics in the Sheepbed Mudstone? 45th
  LPSC Lunar and Planetary Science Conference. 2014\relax
\mciteBstWouldAddEndPuncttrue
\mciteSetBstMidEndSepPunct{\mcitedefaultmidpunct}
{\mcitedefaultendpunct}{\mcitedefaultseppunct}\relax
\EndOfBibitem
\bibitem[Nittler \latin{et~al.}(2020)Nittler, Alexander, Foustoukos, Patzer,
  and Verdier-Paoletti]{nittler2020asuka}
Nittler,~L.; Alexander,~C.; Foustoukos,~D.; Patzer,~A.; Verdier-Paoletti,~M.
  Asuka 12236, The most pristine CM chondrite to date. Lunar and Planetary
  Science Conference. 2020; p 2276\relax
\mciteBstWouldAddEndPuncttrue
\mciteSetBstMidEndSepPunct{\mcitedefaultmidpunct}
{\mcitedefaultendpunct}{\mcitedefaultseppunct}\relax
\EndOfBibitem
\bibitem[Altwegg \latin{et~al.}(2020)Altwegg, Balsiger, H{\"a}nni, Rubin,
  Schuhmann, Schroeder, S{\'e}mon, Wampfler, Berthelier, Briois, \latin{et~al.}
  others]{altwegg2020evidence}
Altwegg,~K.; Balsiger,~H.; H{\"a}nni,~N.; Rubin,~M.; Schuhmann,~M.;
  Schroeder,~I.; S{\'e}mon,~T.; Wampfler,~S.; Berthelier,~J.-J.; Briois,~C.,
  \latin{et~al.}  Evidence of ammonium salts in comet 67P as explanation for
  the nitrogen depletion in cometary comae. \emph{Nat. Astron.} \textbf{2020},
  \emph{4}, 533--540\relax
\mciteBstWouldAddEndPuncttrue
\mciteSetBstMidEndSepPunct{\mcitedefaultmidpunct}
{\mcitedefaultendpunct}{\mcitedefaultseppunct}\relax
\EndOfBibitem
\bibitem[Altwegg \latin{et~al.}(2022)Altwegg, Combi, Fuselier, H{\"a}nni,
  De~Keyser, Mahjoub, M{\"u}ller, Pestoni, Rubin, and
  Wampfler]{altwegg2022abundant}
Altwegg,~K.; Combi,~M.; Fuselier,~S.; H{\"a}nni,~N.; De~Keyser,~J.;
  Mahjoub,~A.; M{\"u}ller,~D.; Pestoni,~B.; Rubin,~M.; Wampfler,~S. Abundant
  ammonium hydrosulphide embedded in cometary dust grains. \emph{Mon. Not. R.
  Astron. Soc.} \textbf{2022}, \emph{516}, 3900--3910\relax
\mciteBstWouldAddEndPuncttrue
\mciteSetBstMidEndSepPunct{\mcitedefaultmidpunct}
{\mcitedefaultendpunct}{\mcitedefaultseppunct}\relax
\EndOfBibitem
\end{mcitethebibliography}

\clearpage

\begin{figure}[H]
\centering
\includegraphics[width=\textwidth]{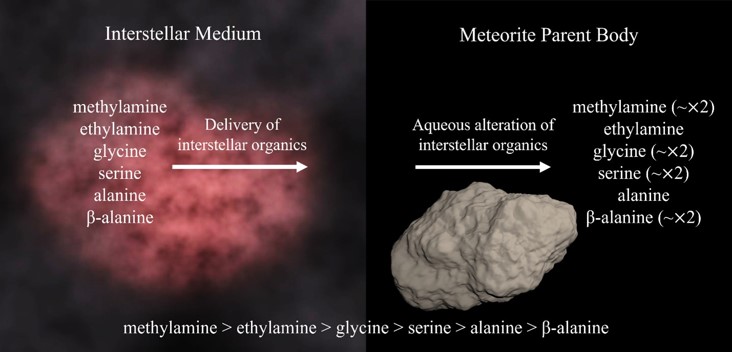}
\caption{For TOC Only} 
\end{figure}

\end{document}